\documentclass{aa}

\usepackage{graphicx}
\usepackage{txfonts}
\usepackage{amsmath}	
\usepackage{amssymb}	
\usepackage[colorlinks, urlcolor = blue, linkcolor = blue, citecolor = blue]{hyperref}
\bibpunct{(}{)}{;}{a}{}{,} 

\usepackage{gensymb}

\begin{document}

   \title{Ejection and Dynamics of Aggregates in the Coma of Comet 67P/Churyumov-Gerasimenko}
   \titlerunning{Ejection and Dynamics of Aggregates in the Coma of Comet 67P}

   \author{P. Lemos\inst{1,2}
          \and
          J. Agarwal\inst{1,2}
          \and
          R. Marschall\inst{3}
          \and
          M. Pfeifer \inst{2}
          }

   \institute{Institut f\"ur Geophysik und Extraterrestrische Physik, Technische Universit\"at Braunschweig, Mendelssohnstraße 3, D-38106 Braunschweig, Germany.\\
              \email{j.lemos-velazquez@tu-braunschweig.de}
         \and
             Max Planck Institute for Solar System Research, Justus-von-Liebig-Weg 3, D-37077 G\"ottingen, Germany.
        \and
            CNRS, Laboratoire J.-L. Lagrange, Observatoire de la Côte d’Azur, Nice, France.
             }

   \date{Received <date>; accepted <date>}
 
  \abstract
   {The process of gas-driven ejection of refractory materials from cometary surfaces continues to pose a challenging question in cometary science. The activity modeling of comet 67P/Churyumov-Gerasimenko, based on data from the Rosetta mission, has significantly enhanced our comprehension of cometary activity. But thermophysical models have difficulties in simultaneously explaining the production rates of various gas species and dust. It has been suggested that different gas species might be responsible for the ejection of refractory material in distinct size ranges. }
   {This work focuses on investigating abundance and the ejection mechanisms of large ($\gtrsim$ 1 cm) aggregates from the comet nucleus. We aim to determine their properties and map the distribution of their source regions across the comet surface. This can place constraints on activity models for comets.}
   {We examined 189 images acquired at five epochs by the OSIRIS/NAC instrument on board the Rosetta spacecraft. Our goal was to identify bright tracks produced by individual aggregates as they traversed the camera field of view. In parallel, we generated synthetic images based on the output of dynamical simulations involving various types of aggregates. By comparing these synthetic images with the observations, we determine the characteristics of the simulated aggregates that most closely resembled the observations.}
   {We identified over 30000 tracks present in the OSIRIS images, derived constraints on the characteristics of the aggregates and mapped their origins on the nucleus surface. The aggregates have an average radius of $\simeq5$ cm, and a bulk density consistent with that of the comet's nucleus. Due to their size, gas drag exerts only a minor influence on their dynamical behavior, so an initial velocity is needed in order to bring them into the camera field of view. The source regions of these aggregates are predominantly located near the boundaries of distinct terrains on the surface.}
   {}

   \keywords{
   methods: data analysis
   -- methods: numerical
   -- comets: individual: \object{67P/Churyumov–Gerasimenko}
           }

   \maketitle

\section{Introduction}\label{sec:intro}

One of the primary scientific objectives of the ESA Rosetta mission was to explore and understand the onset, evolution, and decline of activity exhibited by comet 67P/Churyumov-Gerasimenko (hereafter referred to as 67P) \textit{in situ}. To achieve this goal, the spacecraft was equipped with a suite of complementary instruments. These instruments employed various techniques and targeted different size ranges of particles, allowing for the analysis and collection of data regarding the refractory material found within the coma of 67P. A synthesis of the main findings regarding all size ranges can be found in \citet{Guettler2019}.

While most of dust analysis instruments onboard Rosetta were designed for particles with typical sizes $\lesssim 1$ mm, the mission offered a unique opportunity to observe and study larger particles, exceeding this size range, using the OSIRIS cameras \citep{Keller2007}. The main challenge about using this data comes from the lack of information about their distances and velocities along the line of sight in the images. Several approaches have been taken to address this issue by different authors.

\citet{Fulle2016} used the parallax method, assuming that particles moved radially from the nucleus. Any deviations from this radial path were attributed to spacecraft motion, enabling them to estimate distances. Several authors \citep{Agarwal2016,Pfeifer2022,Pfeifer2024,Shi2024} focused on particles appearing to emerge from the comet limb, treating them as being at the same distance as the nucleus. \citet{Guettler2017} studied particles appearing out of focus and correlated their blurring levels with distances. \citet{Drolshagen2017} and \citet{Ott2017} examined particles detected in both the Wide and Narrow Angle Cameras (WAC and NAC), using the differences in measured position between these cameras to calculate distances. In a different approach, \citet{Frattin2021} utilized images where particles can be seen as bright tracks, estimating distances to the camera based on the length of these tracks and making assumptions about typical particle speeds.

In this study, we employ a method introduced by \citet{Lemos2023}. This method offers an indirect approach based on the comparison between OSIRIS images and synthetic counterparts generated by varying particle properties. The goal is to identify the particle properties that produce the closest matches to the observed images. To achieve this, the method consists on the analysis of OSIRIS images, where particles are visible as tracks due to their relative motion with respect to the spacecraft. The properties of these tracks, namely their orientation angle, length, and integrated brightness, are extracted from the images.

In parallel, dynamical simulations are conducted in order to trace the trajectories of particles. Based on the position of the spacecraft and the camera viewing direction, we generate synthetic images for the times at which the OSIRIS data were obtained. Then, the properties of the tracks within the synthetic images are inferred, and their distributions compared to their real counterparts. The simulated particles generating tracks that best match the observed data are considered to most closely resemble the real particle population.

The primary objective of this study is to conduct an in-depth analysis of various OSIRIS images employing the previously outlined method. This approach enables us to extract valuable insights into the physical characteristics of the aggregates responsible for generating the observed tracks. It also sheds light on the source regions of these aggregates on the surface of the comet and offers an estimation of their flux. This information plays a pivotal role in constraining the activity mechanisms driving aggregate ejections. 

In Section \ref{sec:obs}, we provide a comprehensive overview of the OSIRIS images utilized in our analysis, along with a description of the track detection algorithm. Section \ref{sec:sim} examines the dynamical simulations employed to create synthetic images, which are subsequently compared with the OSIRIS images. Section \ref{sec:res} presents the results derived from this comparison, offering insights into aggregate properties, source regions, and flux. Finally, Section \ref{sec:conclusion} summarizes the key findings of this work.

\section{Observations}\label{sec:obs}

\begin{figure}[t]
    \centering
    \includegraphics[width=\linewidth]{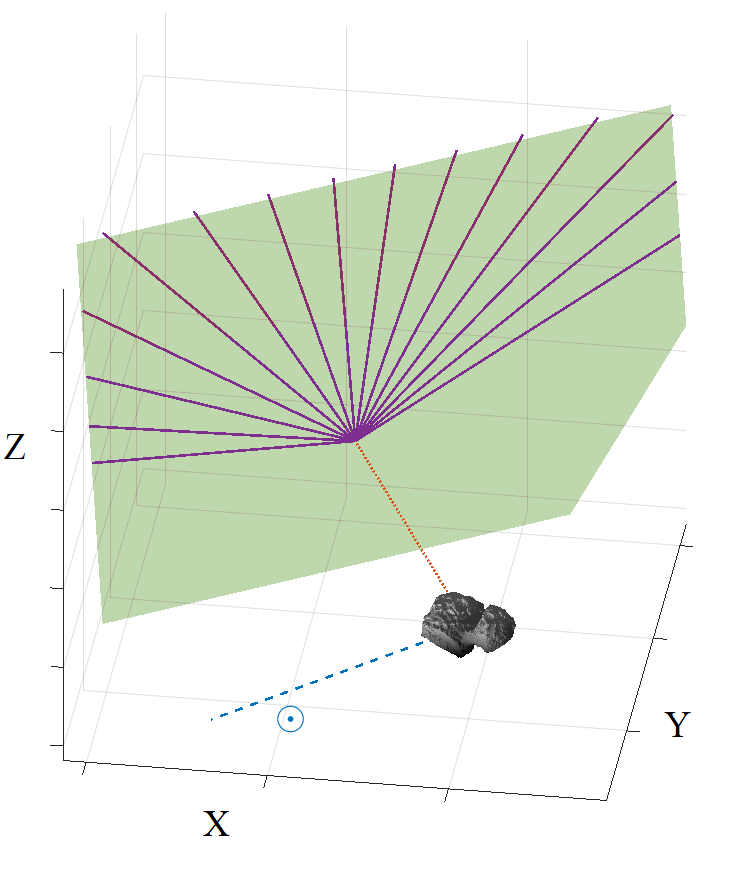}
    \caption{Sketch of the observation geometry for all the image sets used in this work. The blue dashed line indicates the solar direction. The red dotted line indicates the position of the spacecraft. The violet solid lines indicate the different phase angles sampled by the camera, where images with the three different filters were obtained. All these sampled directions belong to the green plane, perpendicular to the nucleus-spacecraft direction. Note that the sizes are not to scale. }
    \label{fig:geom}
\end{figure}

For this work we used images obtained with the OSIRIS NAC. For more details about the characteristics of the camera we refer the reader to \citet{Keller2007}. The images were acquired in five different sets between July 2015 ($r_h=1.32$ au inbound) and February 2016 ($r_h=2.39$ au outbound) with the number of images per set ranging from 21 to 45 (Table \ref{tab:obs}). 
These images were acquired to measure the scattering phase function of the dusty coma, so they share some basic characteristics. The spacecraft was located over the terminator, in such a way that the nucleus-Sun vector was roughly perpendicular to the nucleus-spacecraft vector. The observations sampled different phase angles $\alpha$ on a plane that was perpendicular to the nucleus-spacecraft vector, and for each phase angle three images were taken using different filters. In all the sets used in this work, the images were obtained using the broad-band Blue F24 (480.7 nm), Orange F22 (649.2 nm) and Red F28 (743.7 nm) filters. Figure \ref{fig:geom} shows a sketch of the observation geometry for all sets used in this work.

\begin{table*}
    \caption{Image sets used for this work. Columns represent the mid and short-term planning cycles, date of acquisition, filters used for the acquisition, heliocentric distances, nucleocentric distances and number of images in the set.}
    \label{tab:obs}
    \centering
    \begin{tabular}{c c c c c}
    \hline\hline
    Planning cycle (MTP/STP) & Date & $r_h$ (au) & $r_{S/C}$ (km) & $\#I$\\
    \hline
    018/063 & 2015-07-07 & 1.32 & 153.4 & 45\\
    019/070 & 2015-08-20 & 1.25 & 325.2 & 39\\
    023/086 & 2015-12-14 & 1.89 & 102.6 & 21\\
    025/092 & 2016-01-21 & 2.18 & 79.2 & 39\\
    026/096 & 2016-02-18 & 2.39 & 36.9 & 45\\
    \hline
    \end{tabular}
\end{table*}

These images show a more or less homogeneous background, resulting from the light scattered by small dust grains, as well as bright, straight tracks resulting from the projection onto the image plane of the trajectories of individual aggregates with larger typical sizes on the image plane. While the background brightness in such sequences was studied by, e.g., \citet{Bertini2017} and \citet{Keiser2024} to constrain the spatial distribution of scattering cross-section and the dust scattering properties, we here concentrate on the individual tracks in the foreground, in a data set overlapping with that used by \citet{Bertini2017}.

We analyzed a dataset comprising 189 images. These images were subject to processing, attaining a processing Level 3E according to the OSIRIS scale, equivalent to Level 4 on the CODMAC scale. This implies that the images are solar stray light corrected, in-field stray light corrected, radiometrically calibrated and geometric distortion corrected, expressed in radiance units \citep{Tubiana2015}.

\subsection{Detection method}\label{sec:detMet}

We are interested in analyzing the properties of the aggregates with typical sizes $\gtrsim 1$ cm. As mentioned in Section \ref{sec:intro}, these aggregates are seen as bright tracks in the studied images. We use a semi-automatic method that exploits this property in order to detect these particles. This method, introduced by \citet{Lemos2023} and based on that presented by \citet{Frattin2017}, consists of four automatic steps, plus a last manual one:

\begin{enumerate}
    \item Perform the normalized cross-correlation of the images with track templates in order to generate similarity maps, i.e. a representation of regions in the image that have high probability of containing a track.
    \item Create binary images from the similarity maps.
    \item Detect tracks on the binary images using the Hough transform method \citep{hough1962method,Duda1972}.
    \item Refine the results in order to correct imperfections of the algorithm 
    \item Manually inspect the results.
\end{enumerate}

\begin{figure*}
    \centering
    \includegraphics[width=0.8\linewidth]{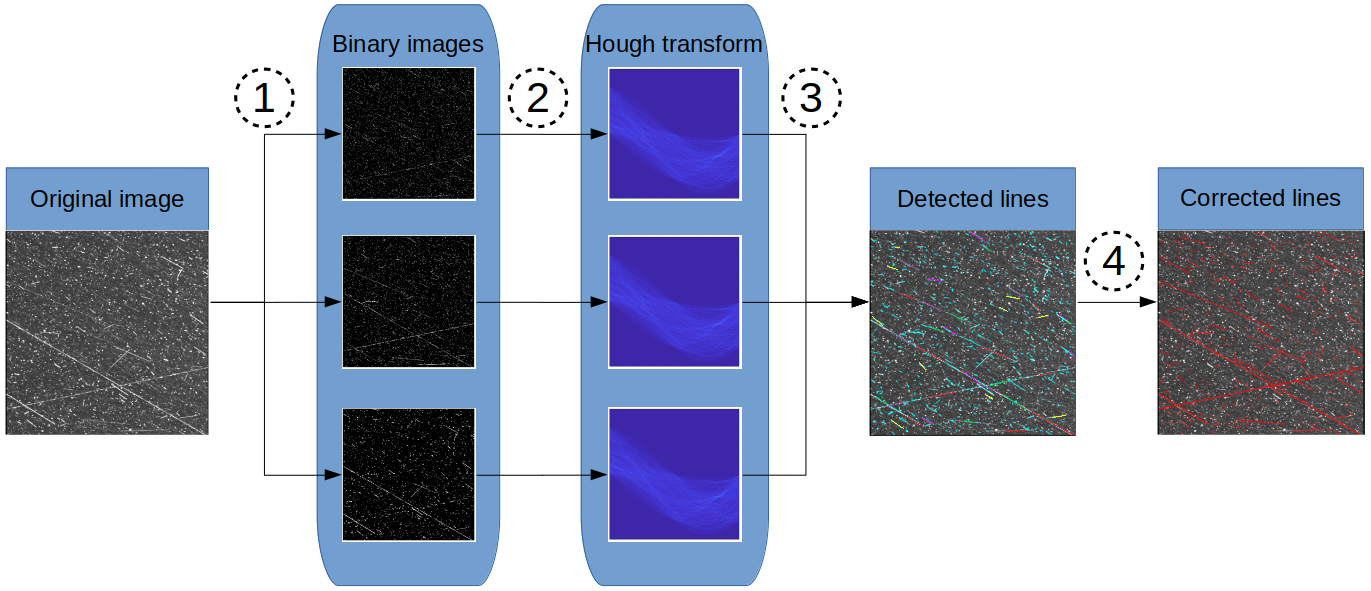}
    \caption{Diagram of the main steps involved in the detection algorithm. }
    \label{fig:algor_diag}
\end{figure*}

A diagram of the automatic steps is presented in Figure \ref{fig:algor_diag}.

By applying this procedure to the groups of images described before, we detected 34616 tracks. The tracks are characterized by three parameters: the orientation angle between the track and the vertical of the image, the total length measured in pixels and the integrated brightness, which was determined using the technique outlined by \citet{Guettler2019}. This method involves performing an aperture photometry analysis for each track, but using a stadium-shaped, rather than a circular aperture. 

\section{Synthetic images}\label{sec:sim}

\subsection{Dynamical simulations}

For an accurate description of the aggregate trajectories, a detailed model of the gas flow in the coma is needed. For this purpose, we use the gas flow results from \citet{Marschall2020}, obtained using a Direct Simulation Monte Carlo (DSMC) method. This code computes the energy input in each facet constituting the nucleus shape model, calculates the surface temperature resulting from the energy balance of incoming light, thermal emission and sublimation, and the gas production rate from each facet, and subsequently computes the steady-state gas distribution in the coma using a DSMC code. In this way, the gas field used for the whole duration of the simulation is the one corresponding to the ejection instant.

The shape model used is a decimated version of the stereo-photogrammetric model SHAP7 \citep{Preusker2015} formed by $\sim440000$ facets with typical length-scales of $\simeq$ 10 m. For more details about this method we refer to \citet{Marschall2016,Marschall2020}.

We used a modified version of the \texttt{DRAG3D} code \citep{Marschall2016,Marschall2020} in order to solve the dust aggregates equations of motion. This code uses a 4th order Runge--Kutta solver with a variable time step to solve the equations describing the motion of spherical dust aggregates in the cometary coma. 

The original code includes the effects of the gravitational and gas drag forces, which we have extended by the solar radiation pressure and solar tidal forces. The gravitational field is computed by discretizing the volume of the mentioned shape model into volume elements with typical sizes $\simeq 30$ m and summing the contributions of each element, assuming an homogeneous density and a total mass $M_N=9.9\times 10^{12}$ kg \citep{Sierks2015}.

The gas drag force $\mathbf{F}_D$ is calculated using the equation

\begin{equation}
    \mathbf{F}_D=\frac{1}{2}C_D\,m_g\,n_g\,\sigma\,|\mathbf{v}_g-\mathbf{v}_d|\,(\mathbf{v}_g-\mathbf{v}_d),
\end{equation}

\noindent where $m_g$, $n_g$ and $\mathbf{v}_g$ are the molecular mass, number density and velocity of the gas respectively, $\mathbf{v}_d$ is the dust velocity and $C_D$ is the drag coefficient. Assuming the gas flow is in equilibrium, $C_D$ is defined as

\begin{equation}
    C_D=\frac{2s^2+1}{\sqrt{\pi}s^3} \mathrm{e}^{-s^2} + \frac{4s^4+4s^2-1}{2s^4}\mathrm{erf}(s) + \frac{2\sqrt{\pi}}{3s} \sqrt{\frac{T_d}{T_g}},
\end{equation}

\noindent where $T_g$ and $T_d$ are the gas and dust temperatures respectively (assumed to be equal) and

\begin{equation}
    s=\frac{|\mathbf{v}_g-\mathbf{v}_d|}{\sqrt{\frac{2k_BT_g}{m_g}}}.
\end{equation}

Solar radiation pressure $\mathbf{F}_R$ and solar tidal forces $\mathbf{F}_T$ are decribed as follows

\begin{align}
    \mathbf{F}_R & = -\frac{C_{\odot} Q_{RP} \pi r_d^2}{r_h^3 c} \mathbf{r}_h\\
    \mathbf{F}_T & = \mathbf{F}_S(\mathbf{r}_C)-\mathbf{F}_S(\mathbf{r}_C+\mathbf{r}_D),
\end{align}

\noindent where $C_{\odot}=1361$ W m\textsuperscript{-2} is the solar constant, $Q_{RP}$ is the radiation pressure efficiency (equal to 1), $c$ is the speed of light, $\mathbf{F}_S(\mathbf{r})$ is the solar gravity force at $\mathbf{r}$, $\mathbf{r}_C$ is the Sun-comet vector and $\mathbf{r}_D$ is the nucleus-dust aggregate vector.

Since the equations of motion are solved in the rotating frame, centrifugal and Coriolis forces are also included in order to account for nucleus rotation. The latter are particularly important for the dust aggregates studied in this work because they have low speed, which implies timescales to traverse the nucleus-spacecraft distance of the order or much longer than the rotation period of the nucleus. For this work we will assume that the dust aggregates do not sublimate volatiles from their surface. For each simulation, a minimum of one and a maximum of ten aggregates are located in a random position of each facet, with the number of particles per facet scaling linearly with the total water production rate (production rate per unit area times area of the facet). The aggregates are given an initial velocity with a direction aligned with the facet normal and a modulus chosen randomly from a Maxwell-Boltzmann (MB) distribution. The MB distribution is parametrized using its most probable velocity $v_P$. With this, the aggregates are characterized by three parameters: particle radius $r_d$, particle density $\rho_d$ and most probable initial velocity $v_P$. The values used for these parameters are shown in Table \ref{tab:dustprop}. The simulations using different combinations of these parameters are carried out independently, and the solutions are later combined in a procedure explained in Section \ref{sec:synthIm}.

\begin{table}
    \caption{Values employed for the aggregate parameter in the dynamic simulations.}
    \label{tab:dustprop}
    \centering
    \begin{tabular}{c c}
    \hline\hline
    Parameter & Values \\
    \hline
    $\rho_d$ & $[100;500;800]$ kg m\textsuperscript{-3}\\
    $r_d$ & $[1; 5; 10; 50]$ cm\\
    $v_P$ & $[0;0.1;0.5;1.0;2.0]$ m s\textsuperscript{-1}\\
    \hline
    \end{tabular}
\end{table}

Since the dust flight time needed to reach the camera FOV is \textit{a priori} unknown, we repeated the dynamic simulations using gas solutions for different sub--solar longitudes spanning a whole comet day in steps of 30\degr. This means that for each simulated epoch, simulations were performed for 12 nucleus rotation states $\times$ 60 aggregate parameters combinations $=$ 720 cases. The total number of aggregates simulated in each case is variable, depending on the distribution of water vapour production on the surface, but is always in the order of $5\times 10^5$ particles. 

\subsection{Generation of synthetic images}\label{sec:synthIm}

To generate synthetic images from which a synthetic track population can be extracted, understanding the motion of aggregates in relation to the spacecraft is crucial. Initially, the position of the spacecraft as well as its observation direction were computed using the SPICE system \citep{Acton1996, Acton2018}. The camera's field of view (FOV) is described as a pyramid with the spacecraft at its apex. Consequently, aggregates observable by the camera are those intersecting this shape. However, due to the characteristics of our simulations such as the high number of particles and the discrete rotation states of the nucleus, determining these intersections is not a straightforward task. Thus, a customized method was devised to address this issue.

At large distances, the gas and dust velocities relative to the nucleus permanently increase \citep{Zakharov2018Icar}. However, it has been shown that the changes of dust velocity above the \textit{acceleration region}, i.e. the region where the gas efficiently accelerates the dust, in the order of 10 km above the nucleus surface depending on the particle size \citep{Gerig2018,Zakharov2018Icar}, are very small. For this reason and in order to simplify the check for intersections between trajectories and the FOV, we assumed that the trajectories of dust aggregates within the range sampled by the camera are straight lines with constant velocity in a non-rotating frame centered on the nucleus. To address potential influences from nucleus gravity and radiation pressure, an additional examination of the trajectories was conducted. The results consistently affirmed that these trajectories indeed followed straight lines, reinforcing the validity of our approximation. 

To ensure a continuous coverage of the entire range of rotation angles, each simulation is deemed valid within the range of $[-15,+15]\degree$ from its nominal rotation angle. The trajectories are parametrized with respect to an angle $\theta_I$ that belongs to that range. Using this parametrization, the minimum distance between the particle trajectory and the line of sight was determined as a function of $\theta_I$. Should the angle between the vector connecting the spacecraft to the aforementioned point of minimal distance and the line of sight direction be smaller than the angular dimension of the FOV ($2.20\degree$), the aggregate in question is said to intersect with the FOV and was then selected for further analysis.

Defining the sides of the FOV as four triangular shapes, we determined the range of angles $\theta_i \in \theta_I$ that provide intersections with each face individually using the algorithm presented by \citet{Moeller1997}. This intersection provides two points, named entry and exit points respectively, which define the direction of the track in the synthetic image. The position of the track endpoints on the synthetic image are associated with the exposure time of the real image. In order to define those endpoints, we selected a random position between entry and exit points. Then, the exposure time of the image was divided in two intervals (not necessarily equal), and the position of this random point is propagated both forward and backward along the trajectory, defining both track endpoints. As we mentioned before, the region observed by the FOV is far away from the aggregate acceleration zone, so it is safe to assume that the aggregate velocity remains constant along their trajectories through the FOV. If an endpoint is not located within the FOV, it is replaced by the corresponding entry or exit point. Lastly, these points are projected onto the image plane, defining the synthetic track.

However, an extra condition must be met by the trajectories in order to accept the synthetic tracks: the time at which particles intersect the FOV must fall within the time range when the corresponding image was obtained. Quantitatively, this is the same as saying that the sub-solar longitude at the moment of observation that is recorded on the image header $\lambda_{obs}$ must be compatible with that of the simulations, obtained by adding the sub-solar longitude at the moment of ejection $\lambda_{ej}$ to the nucleus rotation during the \textit{flight time} $t_F$, defined as the time elapsed from the aggregates ejection to the moment they intersect the FOV. This condition reads as 

\begin{equation}\label{eq:tobs}
    \lambda_{obs} = \lambda_{ej} + \omega_R \times t_F,
\end{equation}

\noindent where $\omega_R$ is the nucleus angular velocity of rotation. 

As mentioned before, $\lambda{ej}$ does not necessarily match with the sub-solar longitude of the gas solutions $\lambda_{gas}$, since the angle $\theta_i$ was added in order to account for discontinuities. Plugging this into Eq. \ref{eq:tobs} we obtain 

\begin{align}
    & \lambda_{obs} = \lambda_{gas} + \theta_i + \omega_R \times t_F\\
    & \theta_i = \lambda_{obs} - \lambda_{gas} - \omega_R \times t_F. \label{eq:theta}
\end{align}

For a trajectory to be accepted, the angle $\theta_i$ found from Equation \ref{eq:theta} has to belong to the range defined by the angles $\theta_I$ that provide intersections. 

For the accepted trajectories we generated synthetic images from the projected tracks, and stored their orientation angle, length and integrated reflectance $B$. The integrated reflectance was found using the equation from \citet{Agarwal2016}

\begin{equation}\label{eq:bright}
    B = \frac{p \Phi(\alpha) I_{\odot} r_d^2}{r_h^2 \Delta^2},
\end{equation}

\noindent where $p$ is the aggregate geometric albedo, $\Phi(\alpha)$ its phase function normalized to 1 at zero phase, $I_{\odot}$ the solar flux in the corresponding filter, $r_d$ is the dust aggregate radius, $r_h$ the heliocentric distance in au and $\Delta$ is the distance between the aggregate and the camera. In case the track was not completely contained inside the synthetic image, we scaled the integrated reflectance $B$ with the fraction of the track belonging to the image. \citet{Lemos2023} found that the phase function of the dust aggregates follow the equation $\Phi(\alpha) = \exp{(-\beta \times \alpha)}$, with a mean $\beta = 8.2\times 10^{-3}$. We used $p=0.065$ \citep{Agarwal2016} and defined $\Delta$ as the mean of the distances between the camera and entry and exit points respectively. 

\subsection{Parameter optimization by comparison between observed and synthetic tracks}\label{sec:inversion}

This comparison was based on the three track properties mentioned in Section \ref{sec:detMet}: orientation angle, length and integrated brightness. However, since the number of tracks detected in different OSIRIS images is highly variable, we generated an estimation of the probability density function (PDF) for the distribution of properties from observed tracks using a Gaussian kernel density estimation. Based on this PDF estimation, we created a new sample of track property triplets which represents the observed ones. Despite the introduction of new statistical uncertainties, this method was preferred since it makes sure that the new set of track properties that will be fitted has a large number of elements for all OSIRIS images studied. 

At this moment, two groups of tracks exist: the ones representing the observed tracks, and the synthetic ones obtained from the dynamical simulations. For each element in the group that represents the observed tracks, we searched the synthetic track pool for the five closest elements in the orientation angle-length-brightness space. This subset of closest neighbours is selected as the result of the comparison procedure. 

\section{Results and discussion}\label{sec:res}

With the procedure presented in previous Sections, we found the properties of the simulated aggregates that provide best fits to the tracks observed in OSIRIS images. In this Section, we will present these properties, as well as the regions on the surface from where these aggregates originate and the mass flux escaping from the comet. Additionally, we will discuss the implications of these results for ejection and activity models. 

\subsection{Effects of gas on particle dynamics}

\begin{figure*}[t]
    \centering
    \includegraphics[width=0.8\linewidth]{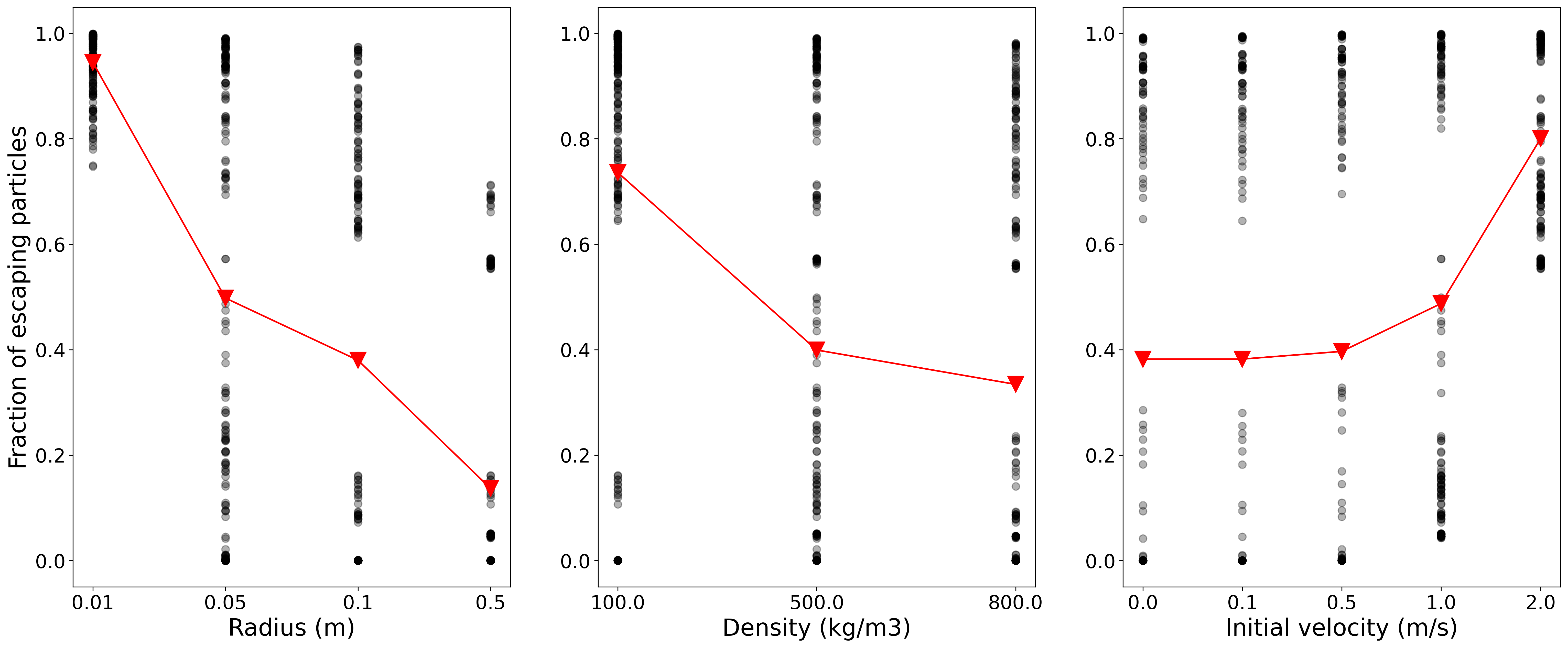}
    \includegraphics[width=0.8\linewidth]{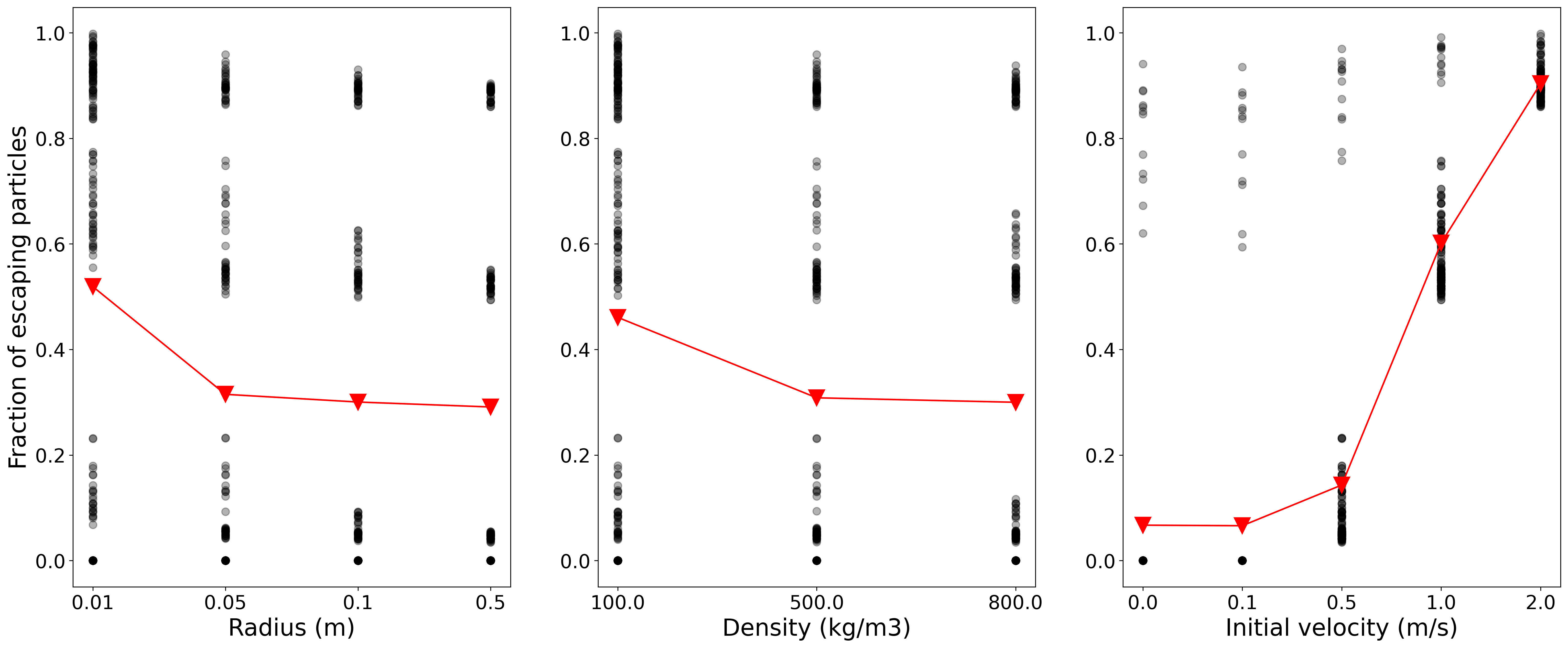}
    \caption{Fraction of aggregates escaping the integration domain for image sets STP070 (top) and STP096 (bottom). Each black dot represents the mean of all particles in a single simulation, whereas the red symbols denote the mean across all aggregates of that particular kind, considering the cumulative data from all simulations.}
    \label{fig:escaping}
\end{figure*}

While simulations without gas were not conducted, we can compare results across sets obtained at different epochs to gain insights into the influence of gas on particle dynamics. Figure \ref{fig:escaping} illustrates the proportion of aggregates escaping the integration domain, categorized based on their aggregate parameters, for the image sets STP070 (7 days after perihelion) and STP096 (189 days after perihelion).

In case of aggregates for which the gas drag forces are highly efficient (small radii and densities), there is a noticeable decrease in the fraction of escaping particles for STP096, which was obtained far from perihelion. In contrast, the fraction of larger, denser aggregates that escape remains relatively constant, indicating a diminished impact of gas on their dynamics.

As expected, the initial velocities also significantly influence the fraction of escaping aggregates. Near perihelion, about 40\% of aggregates with zero initial velocity escape, indicating efficient acceleration by gas drag. Conversely, this number decreases to approximately 5\% for set STP096, highlighting the diminished influence of gas at that point in the orbit.

This change in efficiency highlights the changing nature of the interactions between gas and aggregates phenomenon occurs because we are observing a region where the impact of gas drag shifts from high to low. This influence depends not only on the aggregate properties but also on the characteristics of the gas flow.

\subsection{Aggregate properties}\label{sec:prop}

\begin{figure}[t]
    \centering
    \includegraphics[width=\linewidth]{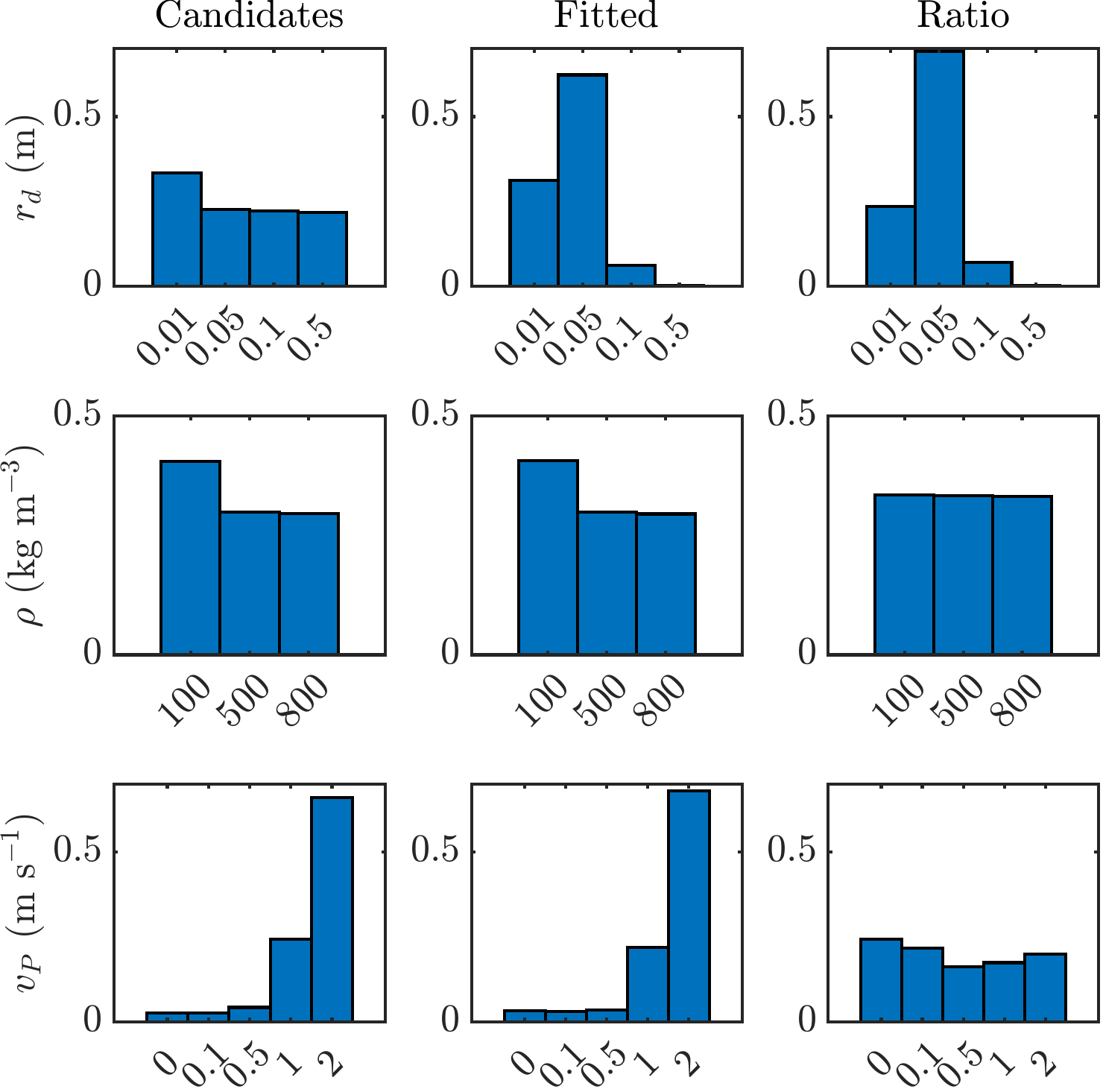}
    \caption{Example of the distribution of physical properties for aggregates in STP092 obtained using our method. From top to bottom, rows show the distribution of radius, density and most probable velocity, while first and second columns show candidates and fitted particles respectively. Last column shows the ratio between heights of columns for fitted aggregates and candidates.}
    \label{fig:propDist}
\end{figure}

We defined two types of simulated aggregates. First, all those aggregates that generate tracks meeting the requirements explained in Section \ref{sec:synthIm} (intersecting the FOV and doing so at a time compatible with the observation) are denominated \textit{candidates}. The subgroup of candidates that generate tracks that are selected as results of the comparison procedure explained in Section \ref{sec:inversion} are called \textit{fitted aggregates}. In order to assess the relevance of each physical property of the aggregates for replicating the observed tracks, we generated histograms for radius, density and most probable initial velocity of candidates and fitted aggregates, considering all the images in each analyzed set. Figure \ref{fig:propDist} shows an example of this type of histograms, in this case for the results of all images from STP092. The histograms for the remaining image sets can be found in Appendix \ref{ap:distProp}.

The data in Figure \ref{fig:propDist} provides insights into the distribution of properties for both candidates and fitted aggregates, with the last column displaying the ratio between the two histograms. The information that can be extracted from them is different. In this particular example, the distribution of candidate radii seems fairly uniform, suggesting that the radius may not be a determining factor for an aggregate to reach the FOV. However, the ratio column reveals a distinct peak around 5 cm, indicating that aggregates of this size are the ones producing tracks similar to the observed ones. In terms of initial velocities, both candidates and fitted particles exhibit similar distributions, emphasizing the importance of this parameter for an aggregate to reach the FOV. Nevertheless, the ratio distribution is approximately uniform, suggesting limited influence on the type of track generated by the aggregates.



\begin{table}[t]
    \caption{Mean of each dust particle parameter considering all the images in the sets.}
    \label{tab:generalPar}
    \centering
    \begin{tabular}{c c c c}
        \hline\hline
        Image set & $\rho_d$ (kg m\textsuperscript{-3}) & $r_d$ (cm) & $v_P$ (m s\textsuperscript{-1})\\
        \hline
        STP063 & 458 & 5.2 & 0.65 \\
        STP070 & 434 & 7.0 & 0.80 \\
        STP086 & 549 & 6.4 & 1.30 \\
        STP092 & 468 & 4.5 & 0.68 \\
        STP096 & 471 & 8.1 & 1.08 \\
        \hline
    \end{tabular}
\end{table}

\begin{figure*}[t]
    \centering
    \includegraphics[width = .8\linewidth]{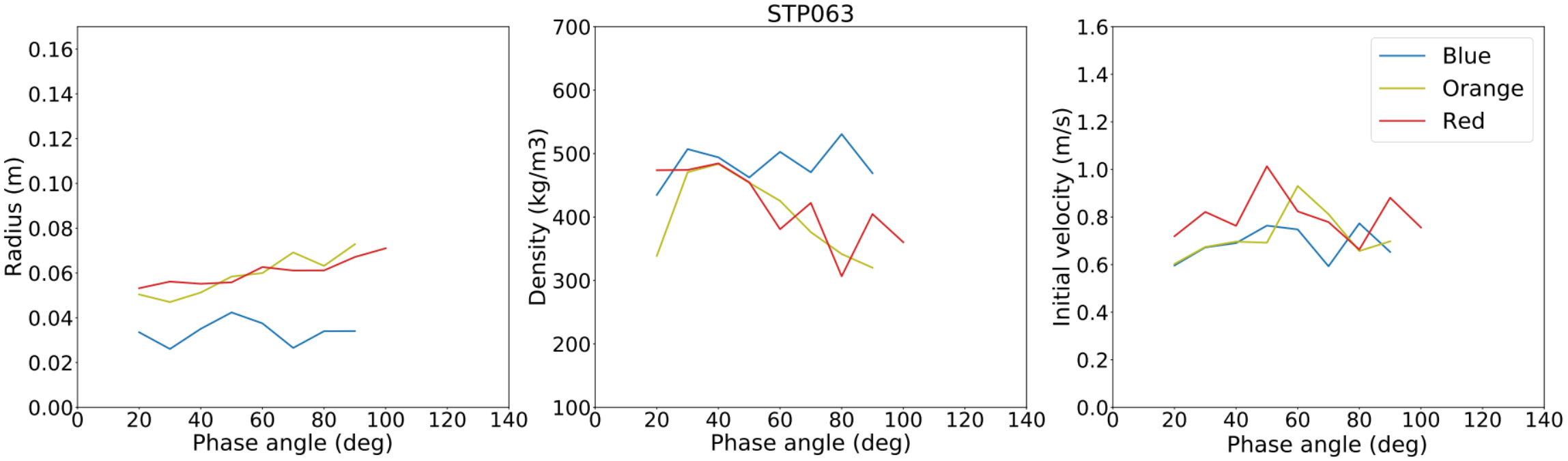}
    \includegraphics[width = .8\linewidth]{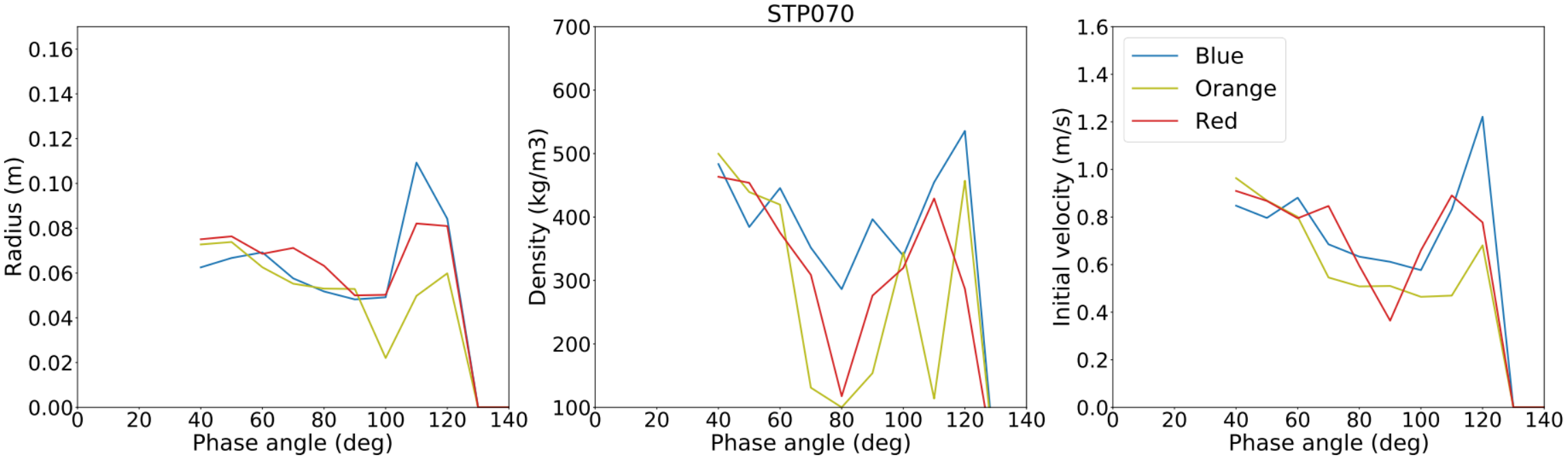}
    \includegraphics[width = .8\linewidth]{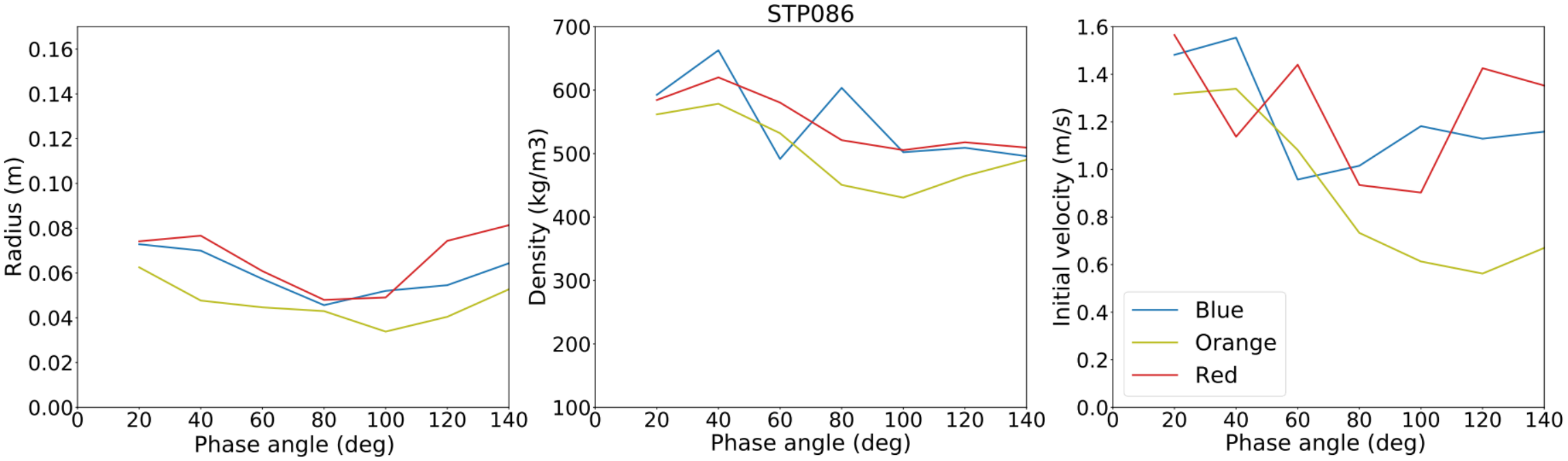}
    \caption{Mean aggregate parameters for sets STP063 (top), STP070 (middle) and STP086 (bottom) as a function of phase angle. Each row represent a different image set, while the columns from left to right represent aggregate radius $r_d$, density $\rho_d$ and initial velocity $v_{in}$ respectively. The colors show the results for images taken with filters Blue F24, Orange F22 and Red F28.}
    \label{fig:dustProp1}
\end{figure*}

\begin{figure*}[t]
    \centering
    \includegraphics[width = .8\linewidth]{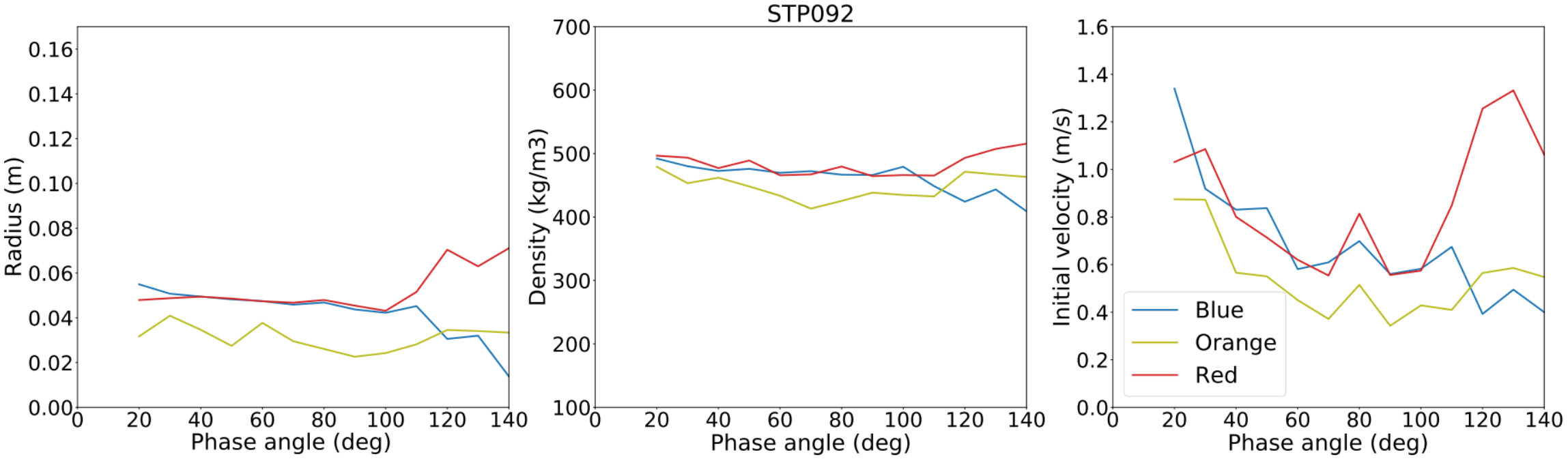}
    \includegraphics[width = .8\linewidth]{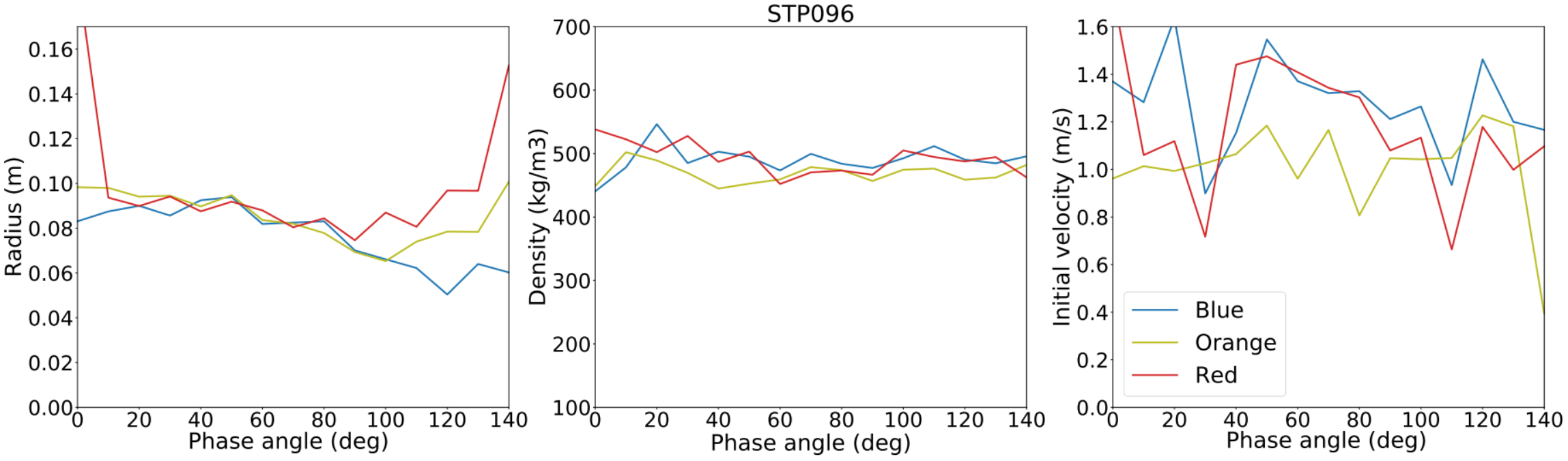}
    \caption{Same as Figure \ref{fig:dustProp1} for image sets STP092 (top) and STP096 (bottom).}
    \label{fig:dustProp2}
\end{figure*}

Figures \ref{fig:dustProp1}-\ref{fig:dustProp2} show the mean value for each physical property for all image sets as a function of phase angle, while Table \ref{tab:generalPar} shows their mean considering all images in each set, averaged across all phase angles\footnote{Note that while the velocity shown in Table \ref{tab:generalPar} represents the most probable initial velocity $v_P$, used to define the MB distribution, Figures \ref{fig:dustProp1}-\ref{fig:dustProp2} represent the initial velocity $v_{in}$ obtained from that distribution.}. Two key observations emerge from these Figures. Firstly, the sets exhibiting greater parameter dispersion are the ones in which the spacecraft was positioned farther from the nucleus. This diminishes the total count of simulated aggregates that reach the FOV, which, in turn, affects the quality of the results, especially at large phase angles. Secondly, the results are similar to those presented in \citet{Lemos2023}, but with a clear difference in particle size. In our case, the mean aggregate sizes lie in the centimeters range, while their results showed aggregates of several decimeters. The source of this difference is a combination of dynamical effects: on the one hand, the model applied here considers additional forces, not taken into account in the mentioned work. On the other hand, the model used for the gas dynamics is different. \citet{Lemos2023} used a fluid model for simulating the gas distribution, while here a DSMC model was used. \citet{Zakharov2018b} showed that the fluid approach yields markedly higher velocity estimates near the nucleus compared to DSMC simulations. Given that these regions exhibit the highest gas densities, the most significant gas drag acceleration occurs there, resulting in significantly greater gas drag expected from the fluid model relative to DSMC. Consequently, according to the predictions of \citet{Lemos2023}, dust aggregate velocities are anticipated to be higher than those obtained here for a given size. To reconcile the discrepancies observed in OSIRIS images, \citet{Lemos2023} adjust these velocities by increasing the sizes of the aggregates, which explains the differences between inferred particle sizes among both approaches.

Although being smaller than the ones found in \citet{Lemos2023}, the aggregate sizes found in this work match well with the ones found by other authors from the analysis of OSIRIS images at a similar range of distances \citep{Agarwal2016,Ott2017,Pfeifer2022,Pfeifer2024}. These aggregates have sizes larger than the ones expected from activity caused by water vapour sublimation, so more volatile species as CO\textsubscript{2} seem to be responsible for their ejection \citep{Gundlach2020}. 

Our results also indicate that the aggregates have a density smaller than the one measured for dust particles using GIADA $\rho_D \simeq 800$ kg m\textsuperscript{-3} \citep{Fulle2016,Fulle2017}, but comparable with the bulk nucleus density $\rho_N = 537.8$ kg m\textsuperscript{-3} \citep{Preusker2017}. This can be interpreted as the aggregates seen by OSIRIS having the same macro-structure as the nucleus rather than that of pebbles. 

\subsection{Initial velocities}

Our method shows that in order to reach the FOV, the aggregates require an initial velocity $v_{in}$ on the order of 1 m s\textsuperscript{-1}. This comes from the fact that, due to their large size, the aggregates are only weakly affected by gas drag, so their initial energy is the main parameter defining whether they are able to reach altitudes sampled by OSIRIS. Most of the work in the field of dust ejection assumes that particles that are initially resting on the surface are lifted by some mechanism, principally gas drag. However, several studies focusing on laboratory measurements, observational data and computational modeling suggest the potential inclusion of an initial velocity component for dust particles. \citet{Yelle2004} and \citet{Huebner2006} showed that in case that dust is ejected from a porous, sub-surface layer of the nucleus, the acceleration provided by gas drag before reaching the surface could be interpreted as an initial velocity, and applied this idea to explain the jets observed in comet 19P/Borrely by the Deep Space 1 spacecraft. Continuing along this line of thought, \citet{Kramer2015,Kramer2016} employed a modeling approach to investigate dust transport and the subsequent formation of structures within the inner coma of 67P assuming a nonzero initial velocity. Examining the coma of comet C/2017 K2, \citet{Kwon2023} conducted an analysis using images obtained by the ground-based Very Large Telescope (VLT). Their findings indicate that the observed structures cannot be adequately accounted for by any of the tested combinations of dust parameters unless an initial velocity is introduced. Laboratory experiments carried out by \citet{Bischoff2019} monitored the trajectories of dust aggregates ejected from a refractory layer situated atop a sublimating water ice block. Using a parabolic function fitting, these authors reported a non-zero initial velocity for the aggregates. 

\begin{figure}
    \centering
    \includegraphics[width=0.75\linewidth]{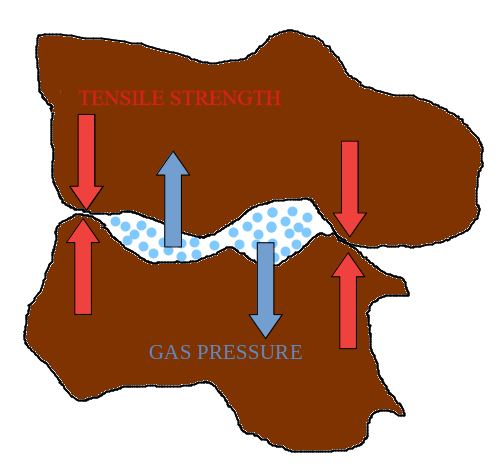}
    \caption{Sketch of the breaking process. Sublimated gas occupies the void space below the aggregate. When pressure equals tensile strength, the aggregate is ejected and a net force coming from the gas pressure acts on it, producing the observed initial velocity.}
    \label{fig:break}
\end{figure}

In this work, we will investigate an analogous explanation to that proposed by \citet{Yelle2004} and \citet{Huebner2006}, which has served as a base for other ejection models, principally dealing with jets (e.g. \citealt{Belton2010,AHearn2011,Vincent2016,Lin2017,Wesolowski2020}). In this model, initial velocity is interpreted as a result of the ejection process itself. Figure \ref{fig:break} provides a schematic representation of this phenomenon, illustrating the scenario where the aggregate is bound to the nucleus, and the void space beneath it is filled with gas. When the gas pressure surpasses the tensile strength, the bond is broken, resulting in a net force on the aggregate. This force acts temporarily, ceasing when the pressurized gas aligns with the state of the surrounding environment.

In a rough approximation, we will assume that all the internal energy of the gas $U_{gas}$ causing the ejection is transferred and converted into kinetic energy of the aggregate $K_{agg}$. It is important to note that in reality, this assumption does not hold true. Thus, this approximation provides an upper limit for the kinetic energy transferred to the aggregate. Assuming that the gas behaves ideally we have

\begin{align}\label{eq:ener}
    K_{agg} & = U_{gas} \nonumber\\
    \frac{1}{2}\,\frac{4}{3} \pi\,r_d^3\,\rho_d\,v_{in}^2 & = C_V \,n\,T,
\end{align}

\noindent where $C_V$ , $n$ and $T$ are the isochoric molar heat capacity, number of moles and temperature of the gas respectively. Since we assumed the gas is ideal, $nT = PV/R$, where $P$ and $V$ are gas pressure and volume respectively, and $R$ is the universal gas constant. With this, Equation \ref{eq:ener} becomes

\begin{equation}\label{eq:vin_part}
    v_{in} = \sqrt{\frac{3\,C_V\,P\,V}{2\,\pi\,r_d^3\,R\,\rho_d}}.
\end{equation}

At the moment of ejection gas pressure equals the material tensile strength $\sigma_T$. While direct measurements of tensile strength within the specified size range are lacking, various measurements across different sizes and theoretical estimations are available. In this study, the focus on large aggregates emphasizes the importance of estimations related to the macro-structure of the comet, as opposed to smaller components like grains or pebbles. For this, we utilize the approximation $\sigma_T = \frac{G}{\sqrt{r_d}}$, where $G = 100$ Pa m$^{1/2}$ \citep{Biele2022}. Lastly, the heat capacity of an ideal gas can be expressed using the equipartition theorem as $C_V = \frac{f}{2} R$, where $f$ represents the number of degrees of freedom of the molecule. Inserting this into Equation \ref{eq:vin_part} yields

\begin{equation}\label{eq:vin}
    v_{in}=\sqrt{\frac{3\,f\,G\,V}{4\,\pi\,\rho_d r_d^{7/2}}}.
\end{equation}

\begin{figure}
    \centering
    \includegraphics[width = .9\linewidth]{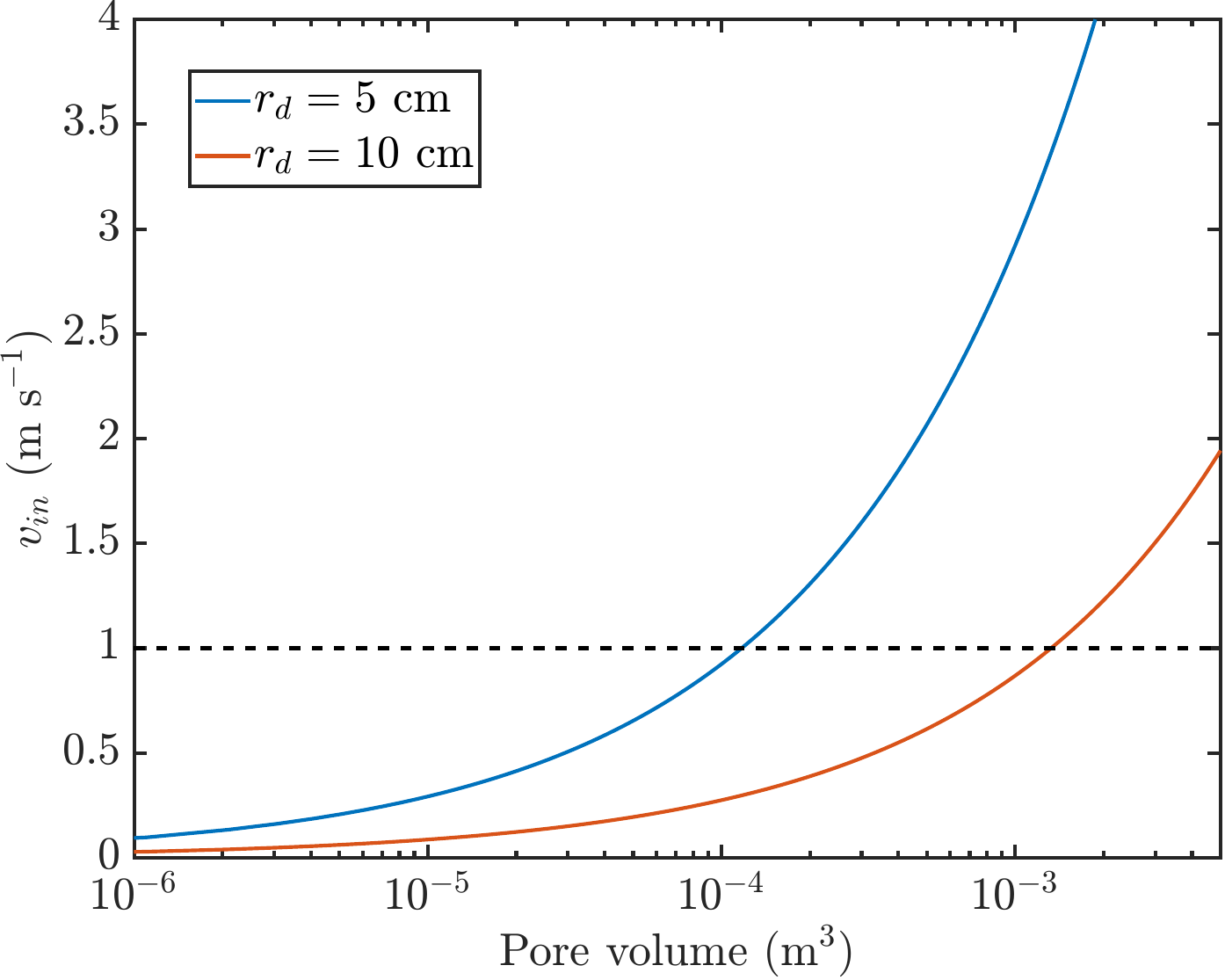}
    \caption{Initial velocity of aggregates as a function of the volume of the pore containing the gas, as determined by Equation \ref{eq:vin}. The blue line corresponds to 5 cm radius aggregates, while the red line indicates 10 cm radius aggregates. The horizontal dashed line indicates an initial velocity of 1 m s\textsuperscript{-1}.}
    \label{fig:vin}
\end{figure}

Since thermophysical models suggest that CO\textsubscript{2} is the most plausible candidate for explaining the ejection of chunks with these sizes. If the vibrational modes are not taken into account, $f$ is equal to 5 for carbon dioxide. Figure \ref{fig:vin} shows the initial velocity as a function of the volume of the pore containing the gas that causes the ejection for aggregates with density $\rho_d$ =500 kg m$^{-3}$ and radii $r_d=$5 and 10 cm. For a 5 cm aggregate, the volume needed to provide a 1 m s\textsuperscript{-1} initial velocity is $1.2\times 10^{-4}$ m\textsuperscript{-3}, while for a 10 cm aggregate this volume increases to $1.3\times 10^{-3}$ m\textsuperscript{-3}. The minimal porosity that provides the necessary energy is expressed by the formula $\phi_{min} = V_{void} / (V_{void} + V_{agg})$, yielding values of approximately 20\% for both scenarios. The macro-porosity of the nucleus was determined to be 0.4 \citep{Fulle2016}, indicating that the pores amid these types of aggregates should contain sufficient gas to generate their acceleration.

\subsection{Source regions}

One advantage of our model is its utilization of a complete shape model, allowing precise identification of the origin locations of the fitted aggregates on the nucleus surface. However, since the flight times are comparable to the nucleus rotation period, minor uncertainties in the determined flight times could significantly impact the determination of the source region, limiting the precision of our method. Therefore, it is essential to note that when referring to source regions, we are indicating a general region rather than pinpointing the exact point from which the aggregates originate.

The trajectory and flight time of the simulated aggregates are influenced not only by the parameters examined in this work, such as gas distribution and velocity, but also by factors not taken into account like particle shape. \citet{Fulle2015} and \citet{Frattin2021} have reported the presence of non-spherical dust particles near 67P, and dynamic analyses have demonstrated that their shape can impact their response to gas drag \citep{Ivanovski2017a,Ivanovski2017b,Moreno2022}. As a result, the findings presented in this section are interpreted in a broad, qualitative manner.

By utilizing the initial positions of candidates and fitted particles used in the dynamic simulations, we connected these positions on the surface to specific facets of the SHAP7 shape model \citep{Preusker2017}, which consists of 125,000 facets. It is important to note that this shape model differs from the one employed in the dynamical simulations, as the SHAP7 model is considered a standard model in comparison. This allowed us to introduce the parameters $C$ and $F$ to represent the counts of candidates and fitted aggregates per facet normalized to their total counts, respectively. In this context, $C$ and $F$ can be interpreted as the probability of a candidate or fitted particle originating from a specific facet. It is important to note that both parameters are constrained within the range of 0 < $F$, $C$ < 1. Combining these parameters, we define the \textit{ejection efficiency} $T$ as 

\begin{equation}
    T = \frac{F-C}{C}.
\end{equation}

The ejection efficiency coefficient can vary within a range starting from $-1$. If the value of $T$ is -1, it signifies that aggregates originating from a particular facet can reach the FOV ($C>0$), but none of them were identified as valid outcomes of the inversion process ($F = 0$). In simpler terms, certain aggregates would be able to generate tracks that are visible in the images, but tracks with those properties were not actually observed. This suggests that the region on the nucleus surface linked with this facet is not effectively ejecting particles of the investigated type. Conversely, facets with $T > 0$ indicate an increase in the count of successfully matched tracks arising from particles originating in that specific surface region.

\begin{figure*}
    \centering
    \includegraphics[width=.8\linewidth]{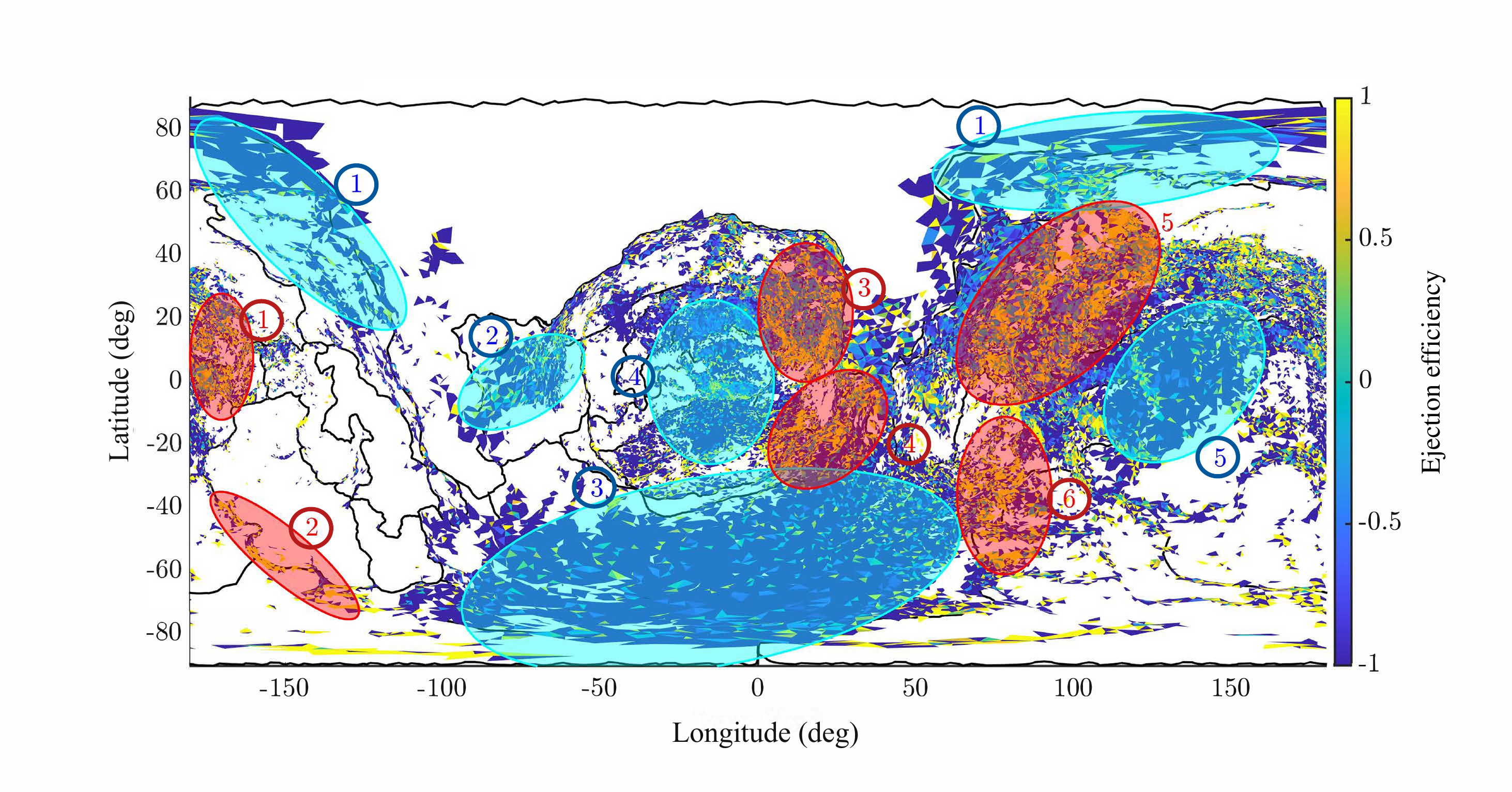}
    \includegraphics[width=0.9\linewidth]{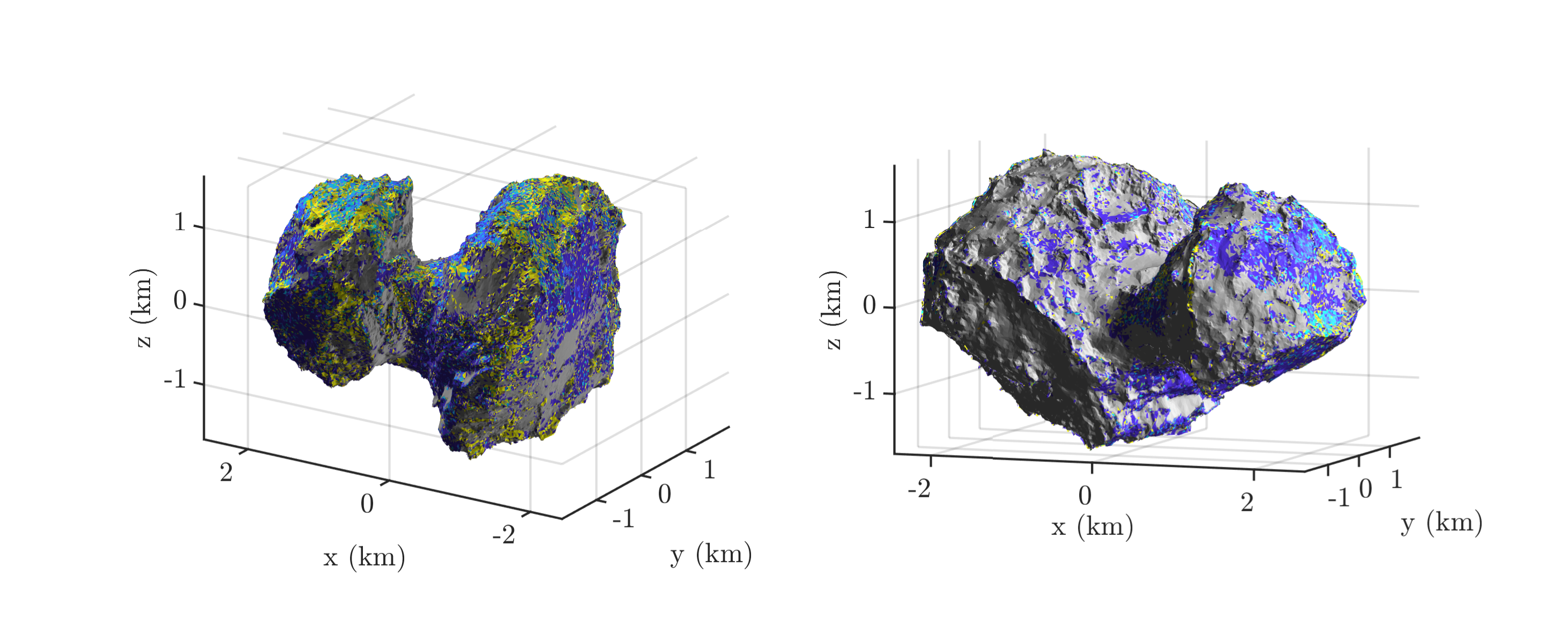}
    \caption{Ejection efficiency for all the sets combined. Blue color indicates patches with low efficiency, while yellow ones show regions with enhanced efficiency. The white and gray areas represent zones which did not provide any candidate in top and bottom panels respectively. Red and blue ellipses show the identified high and low efficiency clusters respectively.}
    \label{fig:spatDist}
\end{figure*}

Figure \ref{fig:spatDist} shows the combined ejection efficiency across all the sets. The colors indicate the value of $T$, while white zones show regions for which no candidates, and hence no fitted particles, could be found. For facets sampled in more than one image set, the mean value was used. The ejection efficiencies for each individual set can be found in Appendix \ref{ap:sourceReg}. We defined zones characterized by high and low ejection efficiency as \textit{clusters}, and identified them in the map shown in Figure \ref{fig:spatDist}. Below, we enumerate these clusters, utilizing the regions defined by \citet{El-Maarry2015,El-Maarry2016} and the sub-regions defined by \citet{Thomas2018}. Additionally, we incorporated the fundamental terrain type, as per the classification by \citet{Thomas2018}, to which each region is attributed.

\medskip\noindent\textbf{High ejection efficiency}

\begin{enumerate}
    \item Ash (Dust) - Apis (Consolidated) - Imhotep (Smooth) - Khonsu (Consolidated): this cluster represents the interface between three very different regions. It comprises the sub-region c of Imhotep up to its boundary with sub-region a, characterized by a intermediate scale roughness and the presence of layers, thought to be exposed by material loss. On the north, Ash-i shows large-scale roughness and layering in cliffs, while Apis is rough in many scales, and similar to the subregion b of Khonsu, one of the most complex regions with evidence of significant change, most probably due to activity \citep{El-Maarry2017}.
    
    \item Bes (Consolidated) - Imhotep (Smooth) - Khonsu (Consolidated): this region runs along the interface between Khonsu-c on the northeast, Bes-a on the south and Imhotep-b on the west. This is a rough area, with several boulders where the surface of Bes drops into Imhotep, forming a layered terrain, and containing a steep cliff that defines the border between Bes and Khonsu.
    
    \item Bastet (Consolidated) - Hathor (Consolidated) - Ma'at (Dust): this is a region that was hard to map all at once due to the change in orientation of the head, where it transitions from north to south hemisphere. It comprises sub-region c of Bastet, a transition zone into the cliff-dominated Hathor, and Ma'at-c, containing many circular depressions that show enhanced activity on a terrain with large-scale roughness. 
    
    \item Aker (Consolidated) - Bastet (Consolidated) - Hapi (Smooth) - Sobek (Consolidated): this region sits mostly on the neck of the comet, where both head and body adjoin the smooth Hapi terrain. Both Aker-d Bastet-c are fractured cliffs, while Sobek-b shows many boulders. 
    
    \item Ash (Brittle) - Aten (Depression) - Babi (Dust) - Khepry (Consolidated): this cluster comprises the boulder-covered depression Aten and all the surrounding sub-regions (dust-covered Ash-a,c,i,j; dusty cliffs Babi-a; rocky, boulder-covered Khepry-a,c). Although formed by diverse terrains, the main characteristic of this cluster is given by the Aten depression, thought to have formed by one or several outburst events \citep{El-Maarry2015}. 
    
    \item Aker (Consolidated) - Anhur (Consolidated) - Bes (Consolidated)-Khepry (Consolidated): this cluster contains the sub-regions a and b from Aker, similar to each other and separated by a ridge. Sub-region a contains tectonic fractures \citep{Thomas2015,Thomas2015b}, while b is similar to Bes and shows a change in slope and roughness when transitioning into Anhur-a. This is a plateau with extreme roughness, containing isolated ridges and pits. On the east it limits with Bes-c, a smooth terrain that shows some boulders from collapsing of cliffs in sub-region d. Lastly, the northeast area is contained in Khepry-a, a flat, rock-like terrain with ponded deposits.  
\end{enumerate}

\textbf{Low ejection efficiency}

\begin{enumerate}
    \item Hapi (Smooth) - Seth (Brittle): both Hapi and Seth are fairly homogeneous regions, without any clear subdivision. The only prominent features are the exposure of very limited areas of consolidated material in Hapi, as well as active pits and semi-circular depressions in Seth. 

    \item Anuket (Consolidated): as in the last case, this region does not have any subdivision. It is smooth at large and intermediate scales, with a rocky appearance at small scale. 
    
    \item Anhur (Consolidated) - Bes (Consolidated) - Geb (Consolidated): although rough at an intermediate scale, the Anhur region seems quite homogeneous, with some cliffs and ridges, principally where it transitions to the neck. The border between Geb and Bes, formed by sub-regions Geb-c, a part of Geb-a and Bes-b, complete this cluster. At this point, Geb mainly shows a cliff, with a smooth transition to Bes. On the western end, Geb-a shows fractured terrain. 
    
    \item Hatmehit (Depression) - Ma'at (Dust) - Wosret (Consolidated): this is a complex cluster, since it shows a very localized high efficiency spot, surrounded by a low efficiency area. The high efficiency spot is located in Hatmehit-b, at the border with Wosret and containing quasi-circular depressions. Around it there are Hatmehit-a, the flat, dust-covered bottom of the depression, and Hatmehit-c, the transition zone to Ma'at, showing a steep cliff with several fractures and layering. On the north, Ma'at-e is covered by dust, but with some of the consolidated material below being exposed, while on the south Wosret-a is located, a flat, smooth surface without any major feature. 
    
    \item Ash (Brittle) - Imhotep (Smooth) - Khepry (Consolidated): this cluster is located mainly in Imhotep-a, with some parts in sub-regions c and d. These are mainly smooth and covered by dust (only sub-region c shows a rougher terrain). Although the adjacent areas Ash-i and Khepry-c show more terrain heterogeneity, this cluster only covers them slightly, so the mentioned areas of Imhotep are the most representatives. 
\end{enumerate}

Initial examination of the morphological properties within the clusters for both high and low ejection efficiencies, reveals no discernible correlation between specific features and the likelihood of aggregate ejections. The sole fundamental terrain type consistently observed in both cluster classes is consolidated, but this observation can be attributed to its prevalence as the most widespread terrain type, encompassing 25.05 km\textsuperscript{2} of the total 51.74 km\textsuperscript{2} surface area (48.5\%).

While high ejection efficiency clusters typically align with regions distant from the axis of rotation, correlating with high centrifugal acceleration, and vice versa, exceptions exist. Notably, low ejection efficiency cluster \#4, situated at the head—where centrifugal acceleration peaks—contradicts this trend. This indicates that while centrifugal acceleration may aid in the ejection of aggregates, it is not the sole determinant of this phenomenon.

Nonetheless, zones of high ejection efficiency tend to be concentrated along the boundaries between regions. These boundaries are marked by the diversity of terrains and the presence of cliffs. Conversely, regions with lower ejection efficiency tend to reside in more uniform areas, typically devoid of prominent features. The source regions of the particles analyzed in this study exhibit a distribution akin to that of dust and gas jets \citep{Vincent2016,Fornasier2019,Lai2019}. These authors attribute strong, sporadic events as the drivers behind jets and outbursts, which can arise through various mechanisms, including cliff collapses, thermal stress fractures, or the pressurization of volatile substances in deeper sub-surface layers.

\subsection{$Af\rho$ and mass loss rate}

In cometary studies, the $Af\rho$ parameter, introduced by \citep{AHearn1984}, is commonly employed to characterize dust activity. Originally designed for ground-based observations, this parameter is calculated as the product of the albedo of cometary dust $A$, the filling factor $f$ defined as the ratio of projected area of dust particles to total sampled area, and the projected aperture radius $\rho$. $Af\rho$ is a valuable parameter for quantifying comet activity because when the ejection and expansion of dust are isotropic and homogeneous, and dust properties remain constant, it is independent of the aperture radius $\rho$.

The derivation of the expression for the $Af\rho$ parameter outlined in this work follows the approach of \citet{Fink2012} and \citet{Fink2015}. While they express $Af\rho$ as a function of the total intensity detected by the sensor, we here define it in relation to the count of particles identified in OSIRIS images.

In its most general form, assuming isotropic ejection, and particles having a constant velocity (stationary coma model), the $Af\rho$ parameter for a single particle size can be expressed as \citep{Fink2012}

\begin{equation}\label{eq:gralAfr}
    Af\rho = \frac{A\,Q_d\,\sigma}{2\,v_d},
\end{equation}

\noindent where $A$ is the dust albedo, $Q_d$ is the production rate of particles of size $r_d$, $\sigma$ is the particle cross section and $v_d$ is the particle velocity. The pivotal component of Equation \ref{eq:gralAfr} is the parameter $Q_d$, so we will now formulate an expression for it using insights gained from our observations and simulations.

In the stationary coma model, the volumetric density of particles can be expressed as

\begin{equation}\label{eq:volDens}
    n_{vol}(r) = \frac{Q_d}{4\pi v_d} \frac{1}{r^2}.
\end{equation}

\begin{figure}
    \centering
    \includegraphics[width = \linewidth]{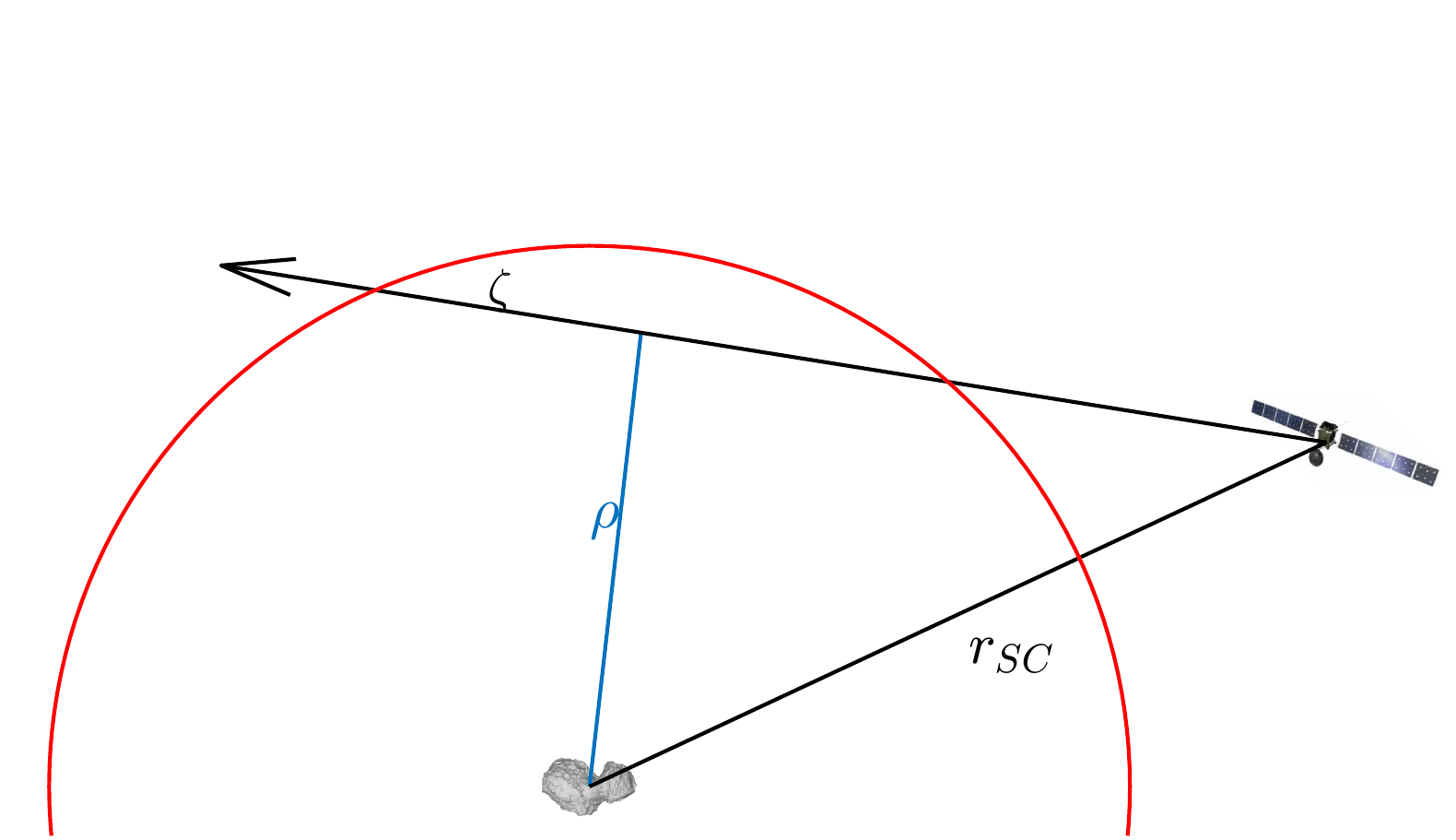}
    \caption{Sketch of the observation geometry for column density determination in the OSIRIS image used in this work.}
    \label{fig:sketch_afrho}
\end{figure}

From this Equation, we derive an expression for the column density as seen by OSIRIS as follows. We define the position of the spacecraft with respect to the nucleus as $\mathbf{r}_{SC}$, while the camera line of sight (LOS) defines a new direction $\mathbf{\zeta}$. The minimum distance between the LOS and the nucleus is defined as the \textit{impact parameter} $\rho$, and the point of minimum distance marks the origin on the $\mathbf{\zeta}$ direction, such that the position of the spacecraft is $-\zeta_{SC}$. Figure \ref{fig:sketch_afrho} shows a sketch of the geometry presented here. 

The column density observed by OSIRIS can be expressed as the integral of the volumetric density over all the observed $\zeta$

\begin{align}\label{eq:ncol}
    n_{col}^{sc} & = \frac{Q_d}{4\pi v_d}\int_{-\zeta_{sc}}^{\zeta_{max}} \frac{1}{\zeta^2+\rho^2} d\zeta \nonumber \\
    & = \frac{Q_d}{4\pi v_d} \frac{\arctan{(\zeta_{max}/\rho)}-\arctan{(-\zeta_{sc}/\rho)}}{\rho}.
\end{align}

As explained in Section \ref{sec:obs}, all the analyzed images were acquired in such a way that the LOS is roughly perpendicular to $\mathbf{r}_{SC}$, so the spacecraft is located at the position $\zeta = 0$ and $\rho$ is the spacecraft altitude. With this, the equivalent ejection rate (the extrapolation of the production rate to the whole nucleus surface based on the particles seen by OSIRIS) can be obtained from Equation \ref{eq:ncol} 

\begin{equation}\label{eq:equivalentDustProf}
    Q_d = \frac{4\,\pi\,v_d\,n_{col}\,\rho}{\arctan(\zeta_{max}/\rho)}.
\end{equation}

Combining Equations \ref{eq:gralAfr} and \ref{eq:equivalentDustProf}, we find the following expression for $Af\rho$

\begin{equation}\label{eq:afrho}
    Af\rho = \frac{2\,\pi\,n_{col}\,\rho\,A\,\sigma}{\arctan(\zeta_{max}/\rho)}.
\end{equation}

\begin{figure}
    \centering
    \includegraphics[width=.8\linewidth]{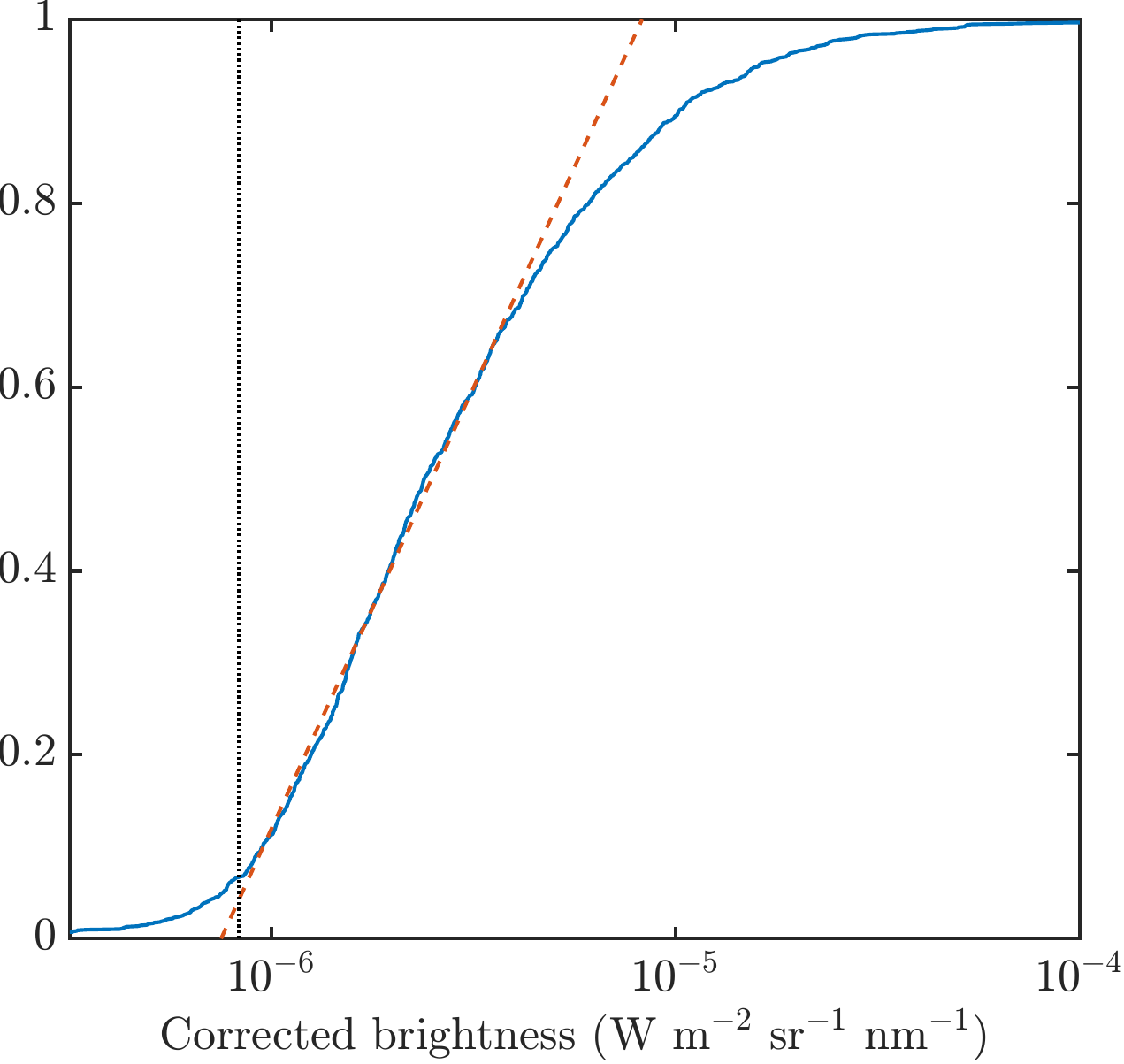}
    \includegraphics[width=.8\linewidth]{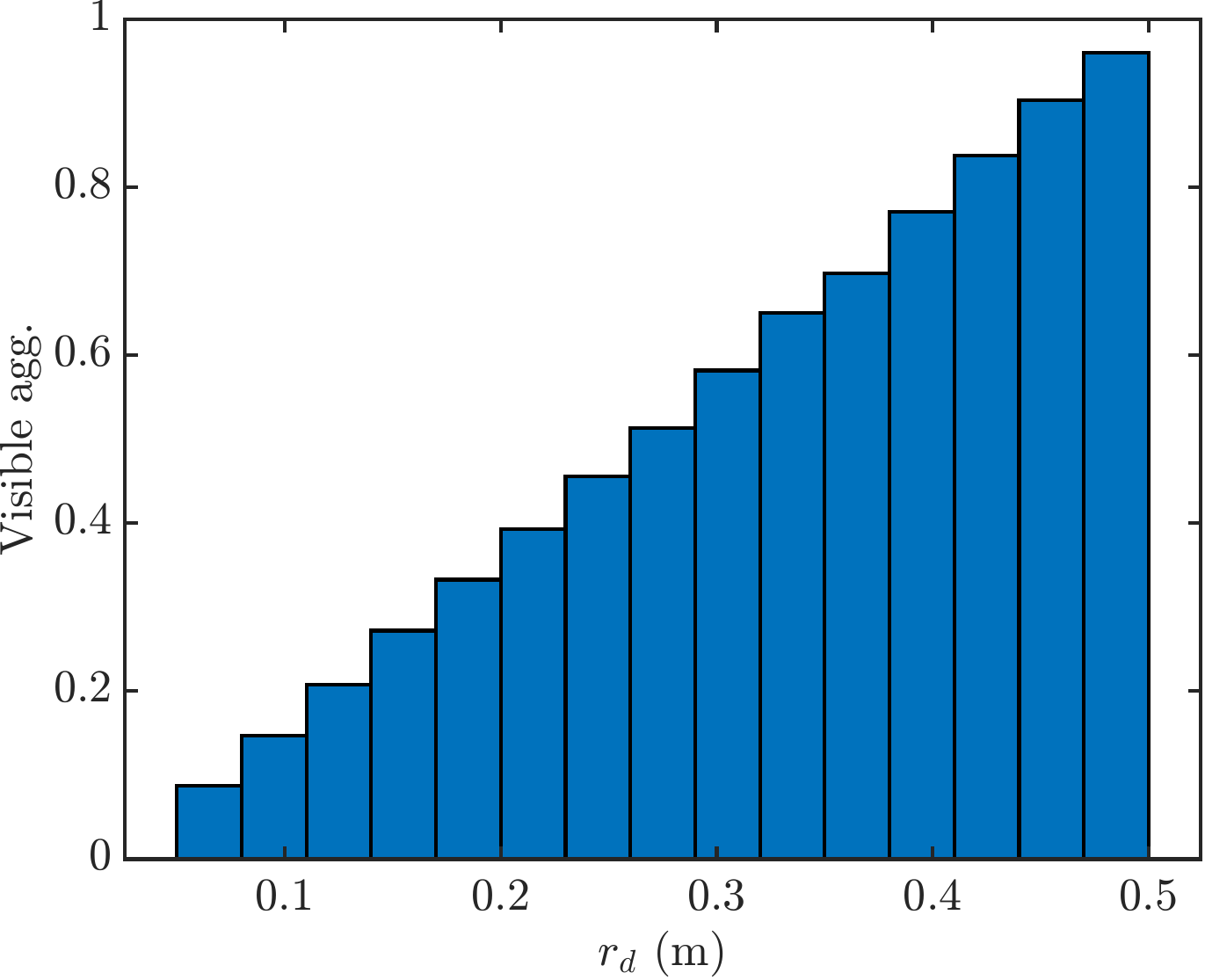}
    \caption{\textit{Top:} Cumulative distribution of phase-function-corrected track brightness for images obtained for set STP096 using the Red filter. The dashed line indicates the best fit for the linear part, while the black dotted line shows the minimum brightness for which the sample is complete. \textit{Bottom:} Fraction of aggregates with brightness larger than $B_{lim}$ at a distance $\leq \zeta_{max}$ from the spacecraft, assuming a stationary coma model.}
    \label{fig:cumBright}
\end{figure}

The primary factor required is the maximum distance $\zeta_{max}$ within which OSIRIS can sample particles. However, the sole information from OSIRIS images refers to the minimum detectable brightness. To define the minimum brightness for which the sample of tracks is complete, we first corrected the measured track brightness to zero phase angle using the phase function mentioned in Section \ref{sec:synthIm}, and then plotted the cumulative distribution of brightness. Figure \ref{fig:cumBright} shows an example of this cumulative distribution for all the tracks detected in OSIRIS images acquired using the Red filter in the set STP092. Using a logarithmic scale for the corrected brightness, these plots exhibit a linear segment within the range of intermediate brightness, followed by a plateau for faint tracks. This plateau highlights the incompleteness of the sample. To address this, we fitted a straight line (in logarithmic space) to the linear segment and determined the limit brightness $B_{lim}$ as the point where the fit separates from the distribution. Figure \ref{fig:cumBright} shows the fit and limit brightness as red dashed and black dotted lines respectively. 

Assuming that the maximum sizes of the aggregates detected in OSIRIS images correspond to the maximum size used for the simulations ($r_d = 50$ cm), we find the maximum distance, $\zeta_{max}$ at which the largest aggregates can be located in order to have a brightness higher than the limit, $B_{lim}$, from Equation \ref{eq:bright}. Nonetheless, it is possible that particles situated closer than $\zeta_{max}$ remain undetected due to their small size. Consequently, an inclusion of a debiasing factor is required to account for particles present within $\zeta_{max}$ but fainter than the limit brightness.

\begin{figure}
    \centering
    \includegraphics[width = .8\linewidth]{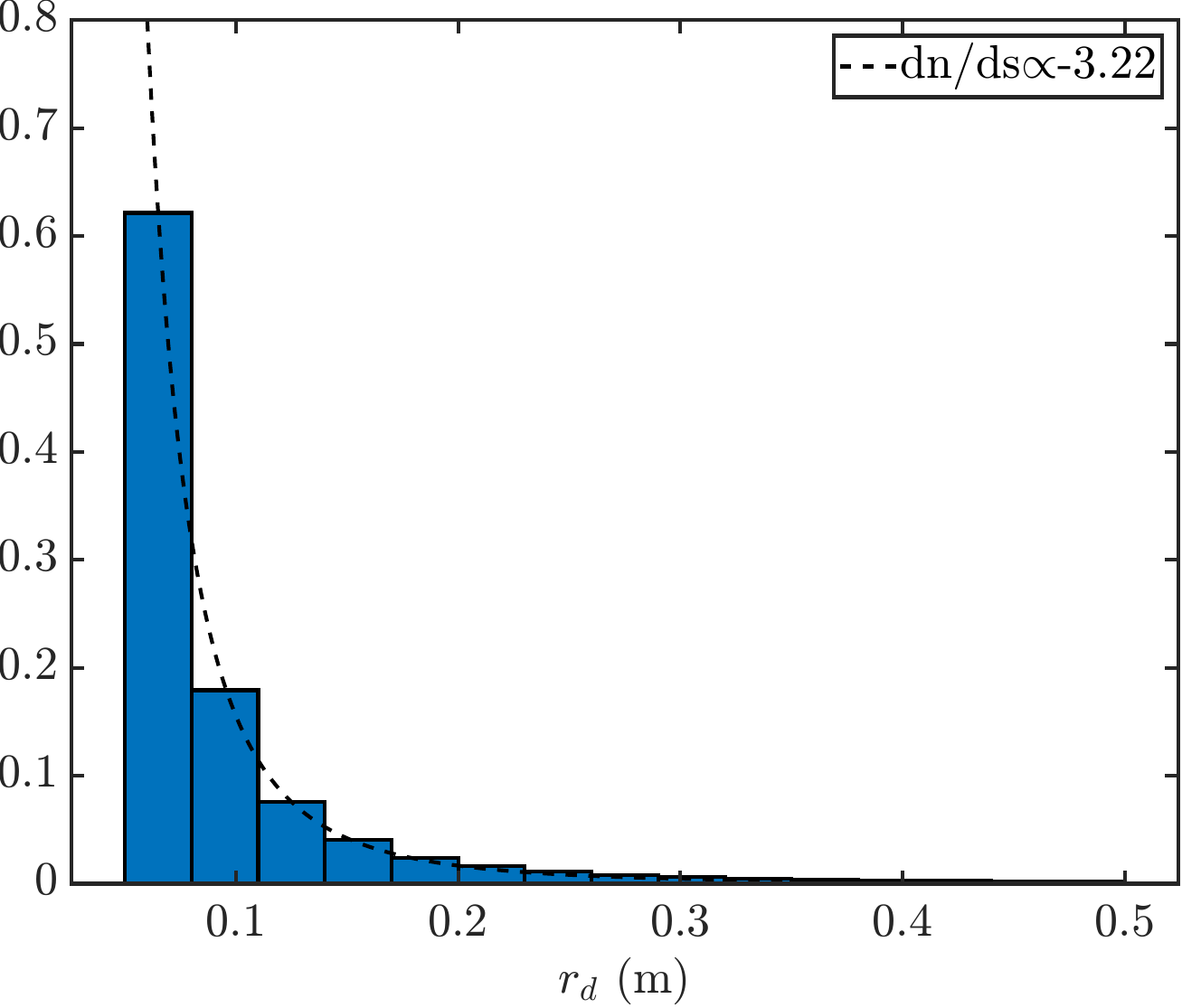}
    \caption{Normalized size distribution obtained from applying the debiasing factor to the size distribution found in \ref{sec:prop}. The best-fit power-law, with a differential index of 3.22, is indicated with a dashed line.}
    \label{fig:sizeDist}
\end{figure}

To determine this debiasing factor, we simulated a set of aggregates under the conditions of the stationary coma model. We calculated the fraction of these aggregates visible to the camera with a brightness larger than $B_{lim}$. This computation yields the aforementioned debiasing factor, which allows to extrapolate the complete size distribution of aggregates ranging from 5 to 50 cm in the vicinity of the spacecraft, which is depicted in Figure \ref{fig:sizeDist}, along with the best-fit power-law for the differential distribution, characterized by an index of -3.22. This value is in agreement with other measurements in the same size range, obtained from different types of measurements, as particles observed by OSIRIS \citep{Agarwal2016,Fulle2016}, boulders on the nucleus surface observed by the lander Philae \citep{Mottola2015} and trail simulations based on ground-based observations \citep{Agarwal2010,Fulle2010}.

\begin{figure}
    \centering
    \includegraphics[width=0.9\linewidth]{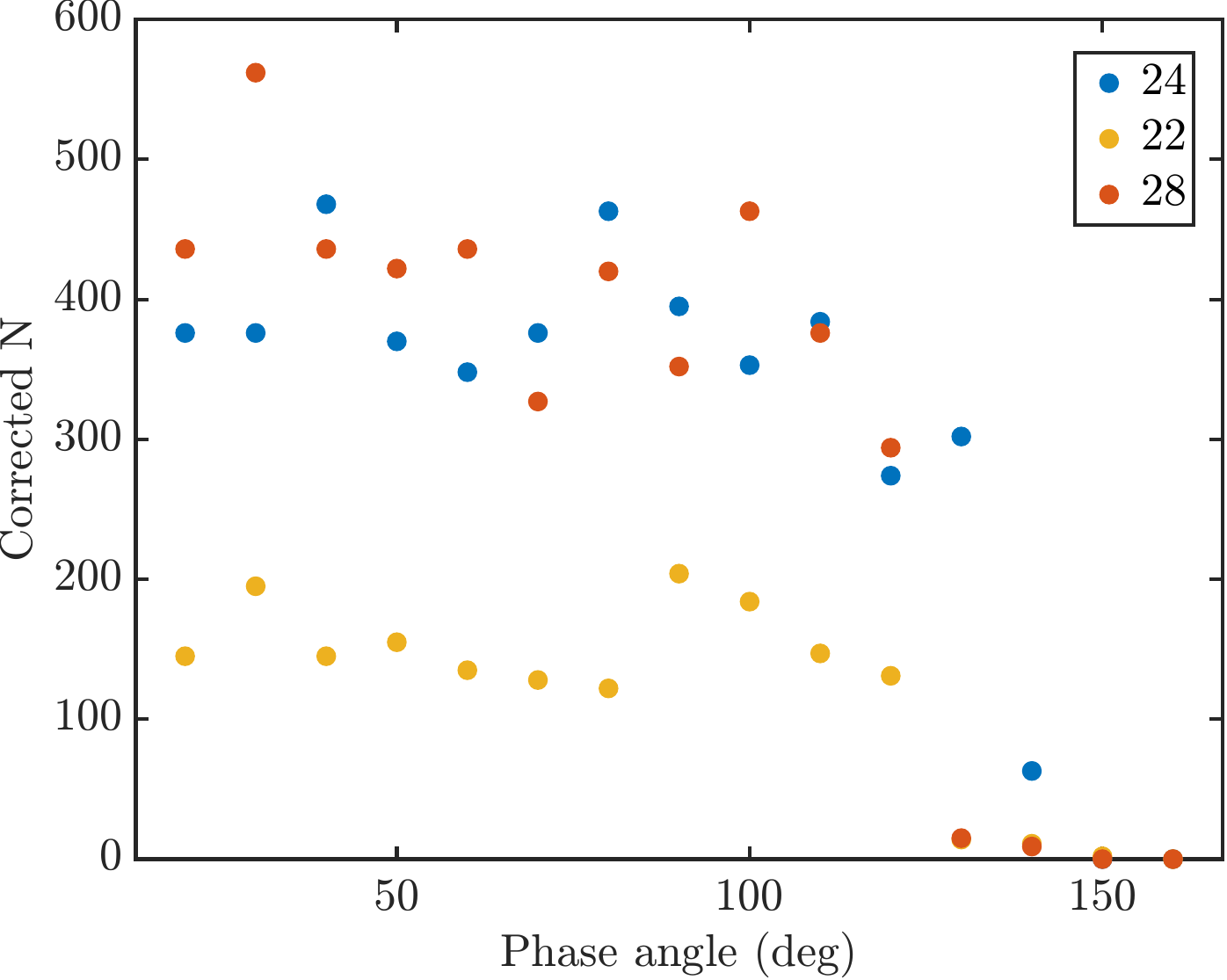}
    \caption{Number of tracks per image after phase-angle-correction and brightness debiasing as a function of the phase angle. This example shows the result for STP063. The colors indicate images obtained using different filters.}
    \label{fig:corrN}
\end{figure}

This debiasing factor, together with the phase correction factor, can be applied to the size distribution extracted from the combination of OSIRIS images and simulations, in order to provide the corrected number of tracks observed in each image (Figure \ref{fig:corrN} and Appendix ). With this procedure, we found the mean number of tracks per image for all the images. We found that images obtained using the Orange F22 filter showed fewer tracks per image, caused by the shorter exposure times used for this type of images compared to the ones acquired with the remaining filters. On the other hand, Blue F24 and Red F28 showed similar numbers of tracks per image, despite the different exposure times used for acquiring images using these filters. This seems to indicate a sort of saturation effect, where the increase of exposure time would not provide a larger number of tracks due to the contribution of light scattered by the diffuse coma. For this reason, the mean track number was calculated using both these filters. 

\begin{figure}
    \centering
    \includegraphics[width=\linewidth]{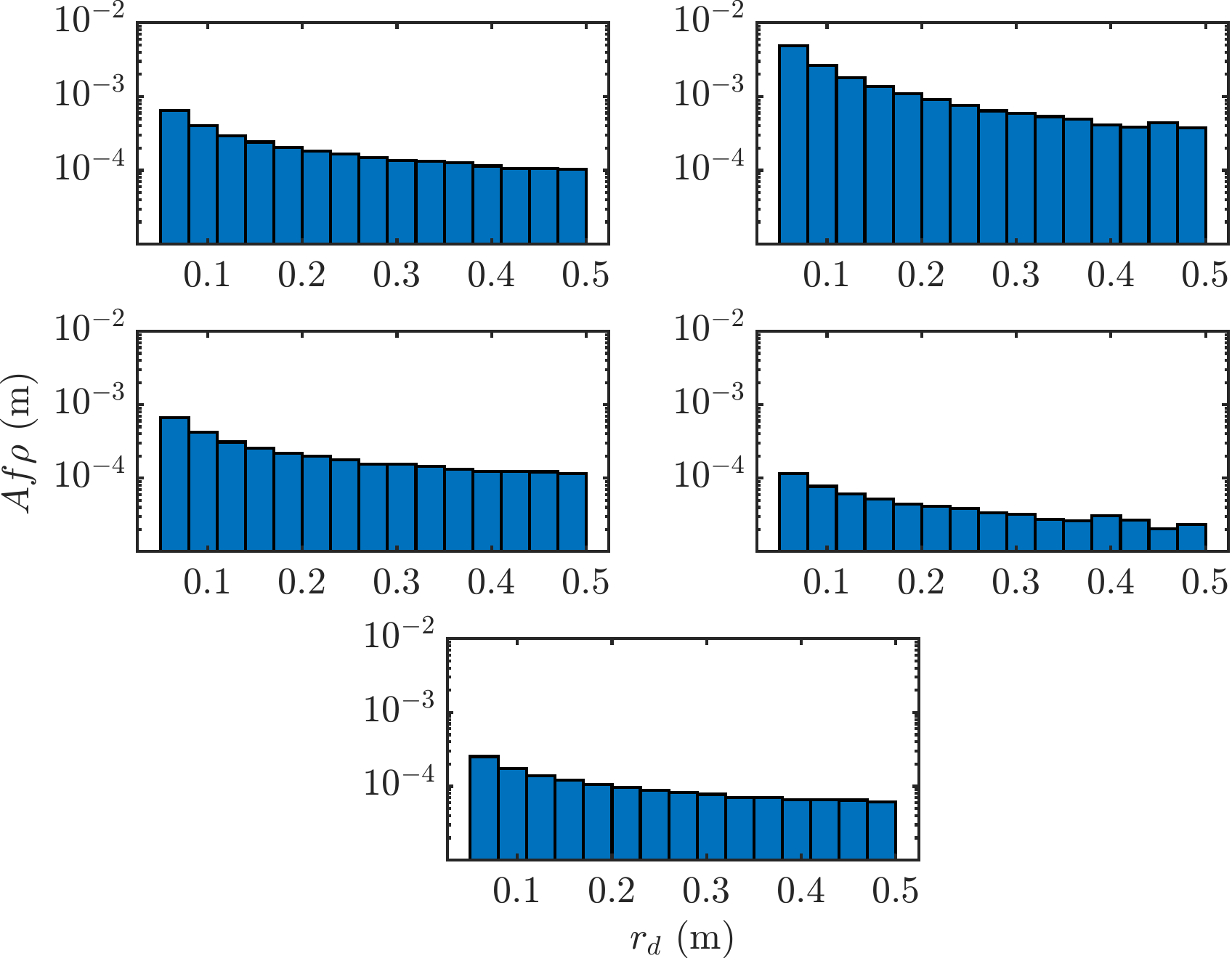}
    \caption{$Af\rho$ parameter for the range 5-50 cm with bin width equal to 3 cm. Starting from top left, the panels show the results for image sets STP063, STP070, STP086, STP092 and STP096 respectively.}
    \label{fig:Afrho}
\end{figure}

Assuming all the particles are located within the distance $\zeta_{max}$ from the spacecraft, the column density can be found as the ratio of the total number of particles and the area base of the pyramid defining the FOV. With all these elements, we calculated the values of $Af\rho$ for each size bin using Equation \ref{eq:afrho}. These results are shown in Figure \ref{fig:Afrho}, while the integrated values of $Af\rho$ for all sizes are presented in Table \ref{tab:afrho}. 

\begin{table*}
    \caption{$Af\rho$, particle flux $Q_n$ and mass flux $Q_m$ integrated for all particle sizes in the range 5-50 cm.}
    \label{tab:afrho}
    \centering
    \begin{tabular}{c c c c c}
        \hline
        \hline
        Image set & Days from perihelion & $Af\rho$ (cm) & $Q_n$ (s\textsuperscript{-1}) & $Q_m$ (kg s\textsuperscript{-1}) \\
        \hline
        STP063 & -37 & 0.31 & 2.44 & 12.9\\
        STP070 & 7 & 1.73 & 13.27 & 49.5\\
        STP086 & 123 & 0.33 & 1.78 & 8.2 \\
        STP092 & 161 & 0.07 & 0.28 & 1.6 \\
        STP096 & 189 & 0.15 & 0.84 & 7.2 \\
        \hline
    \end{tabular}
\end{table*}

Our $Af\rho$-values lie in the mm-cm-range and are considerably smaller than the typical values encountered for 67P derived from ground-based telescope images \citep{Boehnhardt2016,Snodgrass2017} or the diffuse coma as seen by Rosetta \citep{Rinaldi2016,Gerig2018}, which primarily fall in the meter scale. This outcome is expected, given that these studies predominantly focus on smaller dust particles, which are anticipated to be more abundant. However, also the $Af\rho$ values presented by \citet{Fulle2016b} and \citet{Ott2017} for a similar size range as investigated here are several orders of magnitude larger than ours. A possible cause of the discrepancy may lie their employing much shorter distances between particles and camera (below 10 km, compared to $\zeta_{max}\simeq 60$ km here), which significantly augments the particle density.

\citet{Fulle2016b} finds that the aggregate loss rate can be found through the expression 

\begin{equation}
    Q_n = \frac{(Af\rho)_b\,v_d}{2\,A_p\,\sigma},
\end{equation}

\noindent where $(Af\rho)_b$ represents the contribution to the cumulative $Af\rho$ originating from a particular bin, $v_d$ denotes the average particle velocity, that can be determined from our dynamical simulations and $A_p=0.065$ is the geometric albedo at 649 nm \citep{Fornasier2015}. The mass loss rate is straightforwardly defined as $Q_m = Q_n \times m_b$, where $m_b$ represents the characteristic mass corresponding to each bin. Table \ref{tab:afrho} shows the total particle and mass fluxes (integrated over all the bins) for the five image sets analyzed in this work.

\begin{figure}
    \centering
    \includegraphics[width=0.8\linewidth]{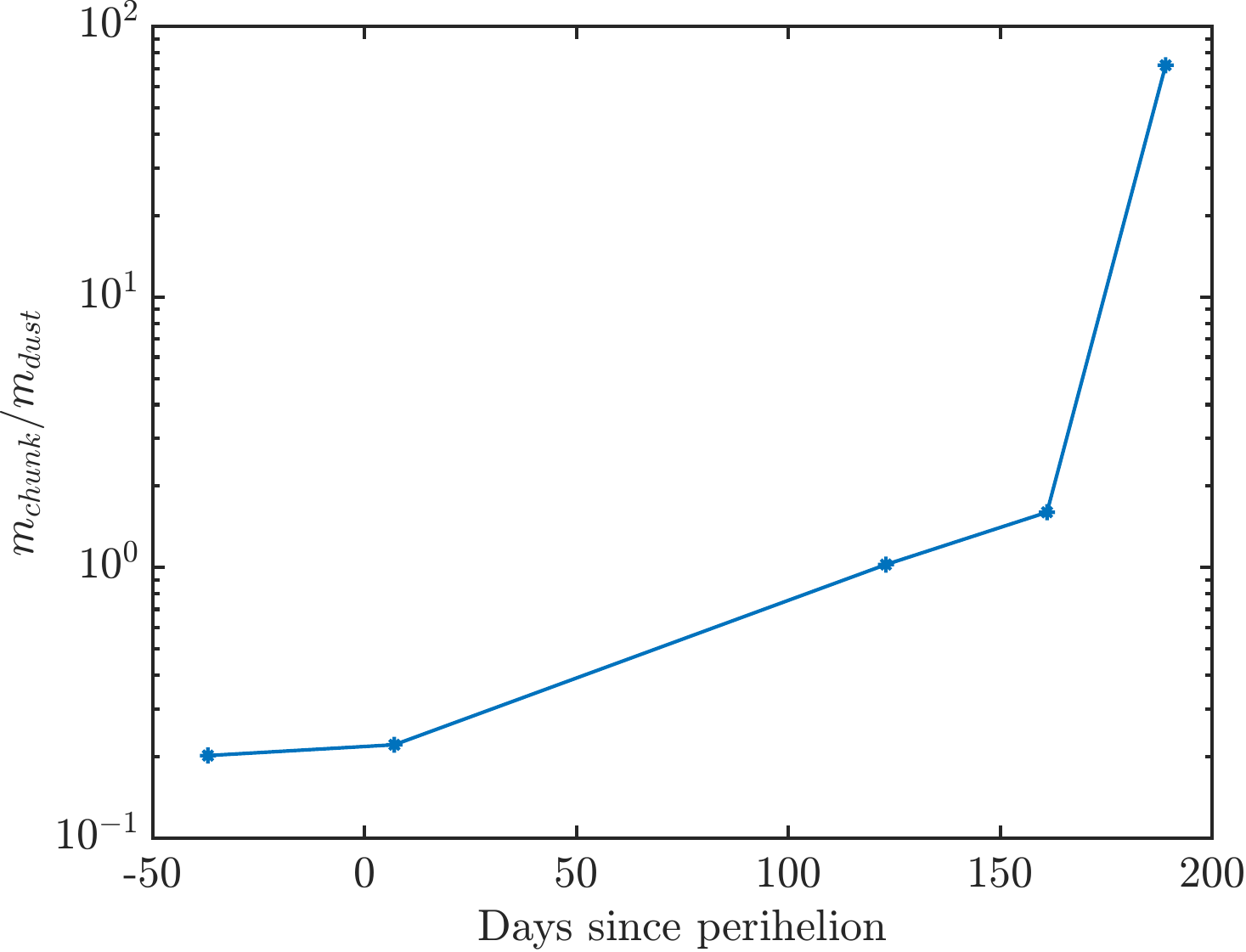}
    \caption{Ratio of mass lost in large aggregates to that lost in small dust. The mass loss rate for dust dominating the diffuse coma brightness (predominantly small dust) is taken from \citet{Marschall2020}.}
    \label{fig:ratioMassLoss}
\end{figure}

Since the mass loss rate found here accounts for large aggregates, we can compare these values to the mass lost in smaller dust. In their work, \citet{Marschall2020} calculate the mass loss rate for a dust size distribution with a minimum size $r_{min} = 0.1$ $\mu$m by comparing the measured diffuse coma brightness with the one obtained from numerical simulations, method more sensitive to small dust. By comparing their values with ours, we determine the ratio of mass lost in large chunks to that lost in fine dust. These results are shown in Figure \ref{fig:ratioMassLoss}. 

Close to perihelion, the mass lost in large aggregates represents only for around 20\% of the total refractory mass lost. However, as the comet recedes in its orbit, this ratio increases up to values $\sim 100$. A comparable trend in the ratio of water and carbon dioxide production rates was observed by \citet{Laeuter2020}. This finding supports the hypothesis of activity in these size ranges being driven by distinct species: while small grains would be ejected by water sublimation, ejection of large chunks would be driven by CO\textsubscript{2} sublimation. In proximity to perihelion, elevated temperatures lead to a substantially greater water production rate. As the comet moves away from the Sun, the decrease in CO\textsubscript{2} production rate is less steep compared to that of water, explaining the observed change in mass loss rates.

\section{Conclusions}\label{sec:conclusion}

We examined tracks from 189 images acquired by OSIRIS and organized into five sets, each obtained at a different epoch. Through dynamical simulations, we generated synthetic images for a wide range of dust parameters. By identifying the closest matches between simulation and observation, we deduced properties of the aggregates generating the tracks and gained insight into the ejection mechanism.

Our main findings are listed below.

\begin{itemize}
    \item The aggregates have typical mean sizes of $5-10$ cm, bulk densities roughly matching that of the nucleus, and initial velocities of order 1 m s\textsuperscript{-1}.
    \item Although showing some variation, these values are consistent throughout all the analyzed time period, ranging from 37 days before perihelion to 189 after it. 
    \item The proposed ejection mechanism, namely CO\textsubscript{2} sublimating from inner layers and overcoming tensile strength, would be able to provide enough energy to the aggregate in order to reach the required initial velocity.
    \item Likely source regions on the surface correspond to boundaries between morphological regions, marked by a heterogeneity of terrain. Conversely, zones with low aggregate ejection efficiency concentrate on smooth, heterogeneous regions. The source regions found here have similar characteristics to those found for jets by \citet{Vincent2016} and \citet{Fornasier2019}.
    \item The $Af\rho$ parameter contributed by aggregates between 5 and 50 centimeter falls approximately in the $0.1-1$ cm range for all the studied period, much smaller than the contribution by fine dust.
    \item The ratio of mass loss rate in large aggregates versus in fine dust increases with heliocentric distance, supporting the hypothesis that while small grains are ejected by water activity, CO\textsubscript{2} is responsible for activity of large chunks.
\end{itemize} 

The database, containing information from a grand total of 34616 tracks, is available online for public consultation \citep{data}. 

\begin{acknowledgements}
    We thank the anonymous referee for their valuable feedback. We also thank Nick Attree, Manuela Lippi and Johannes Markkanen for the helpful discussions. OSIRIS was built by a consortium of the Max-Planck-Institut für Sonnensystemforschung, Göttingen, Germany; the CISAS University of Padova, Italy; the Laboratoire d’Astrophysique de Marseille, France; the Instituto de Astrofísica de Andalucia, CSIC, Granada, Spain; the Research and Scientific Support Department of the European Space Agency Noordwijk, The Netherlands; the Instituto Nacional de Técnica Aeroespacial, Madrid, Spain; the Universidad Politécnica de Madrid, Spain; the Department of Physics and Astronomy of Uppsala University, Sweden; and the Institut für Datentechnik und Kommunikationsnetze der Technischen Universität Braunschweig, Germany. The support of the national funding agencies of Germany (DLR), France (CNES), Italy (ASI), Spain (MEC), Sweden (SNSB), and the ESA Technical Directorate is gratefully acknowledged. We thank the Rosetta Science Ground Segment at ESAC, the Rosetta Missions Operations Centre at ESOC and the Rosetta Project at ESTEC for their outstanding work enabling the science return of the Rosetta Mission. PL, JA and MP acknowledge funding by the ERC Starting Grant No. 757390 Comet and Asteroid Re-Shaping through Activity (CAstRA). PL and MP conducted the work in this paper in the framework of the International Max-Planck Research School (IMPRS) for Solar System Science at the University of Göttingen. JA acknowledges funding by the Volkswagen Foundation. 
\end{acknowledgements}

\bibliographystyle{aa}
\bibliography{main}

\begin{appendix}
    \section{Distribution of physical properties}\label{ap:distProp}

    \begin{figure}[ht]
        \centering
        \includegraphics[width=\linewidth]{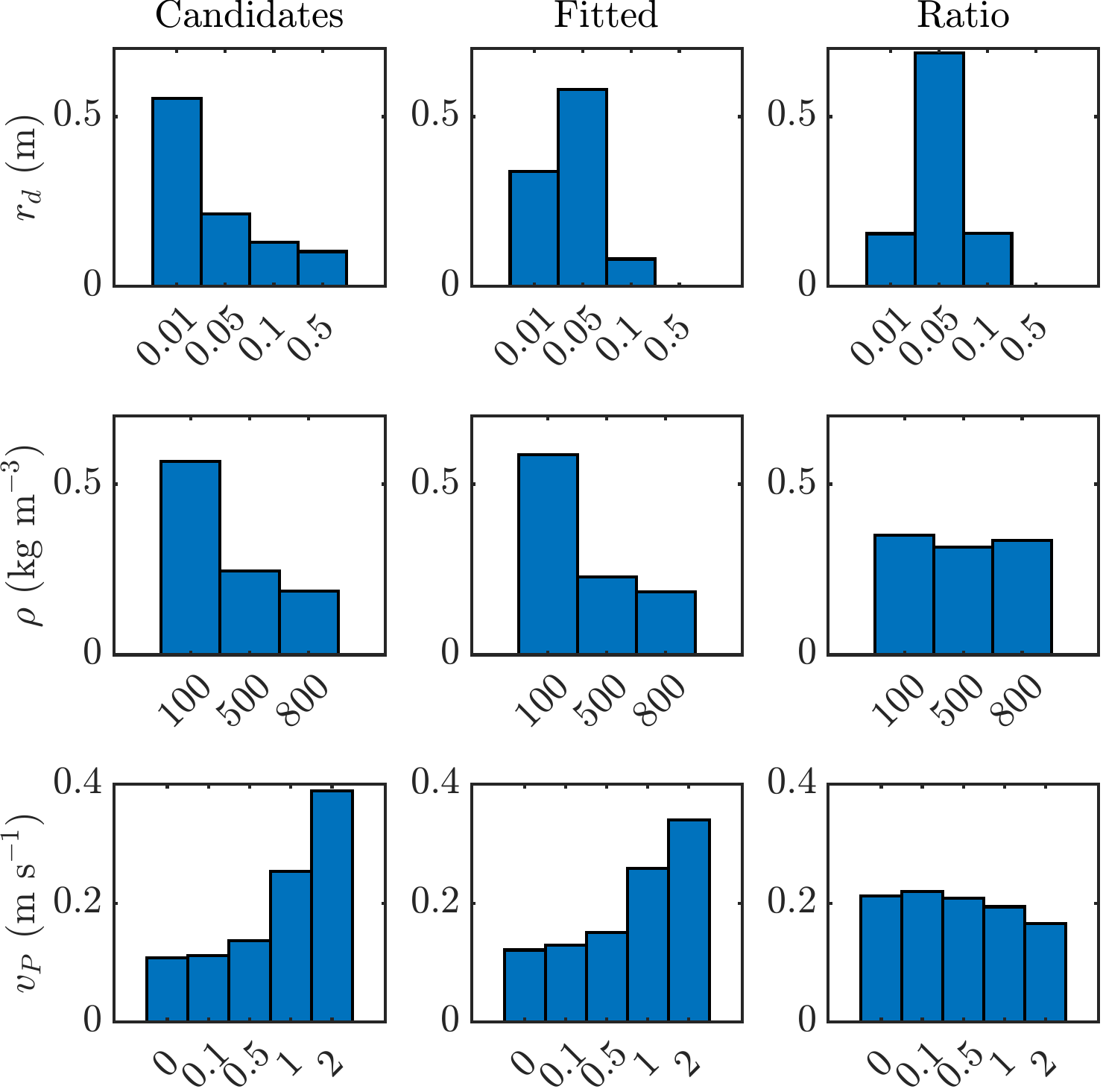}
        \caption{Distribution of physical properties of the aggregates for STP063.}
        \label{fig:distProp_63}
    \end{figure}
    \begin{figure}[ht]
        \centering
        \includegraphics[width=\linewidth]{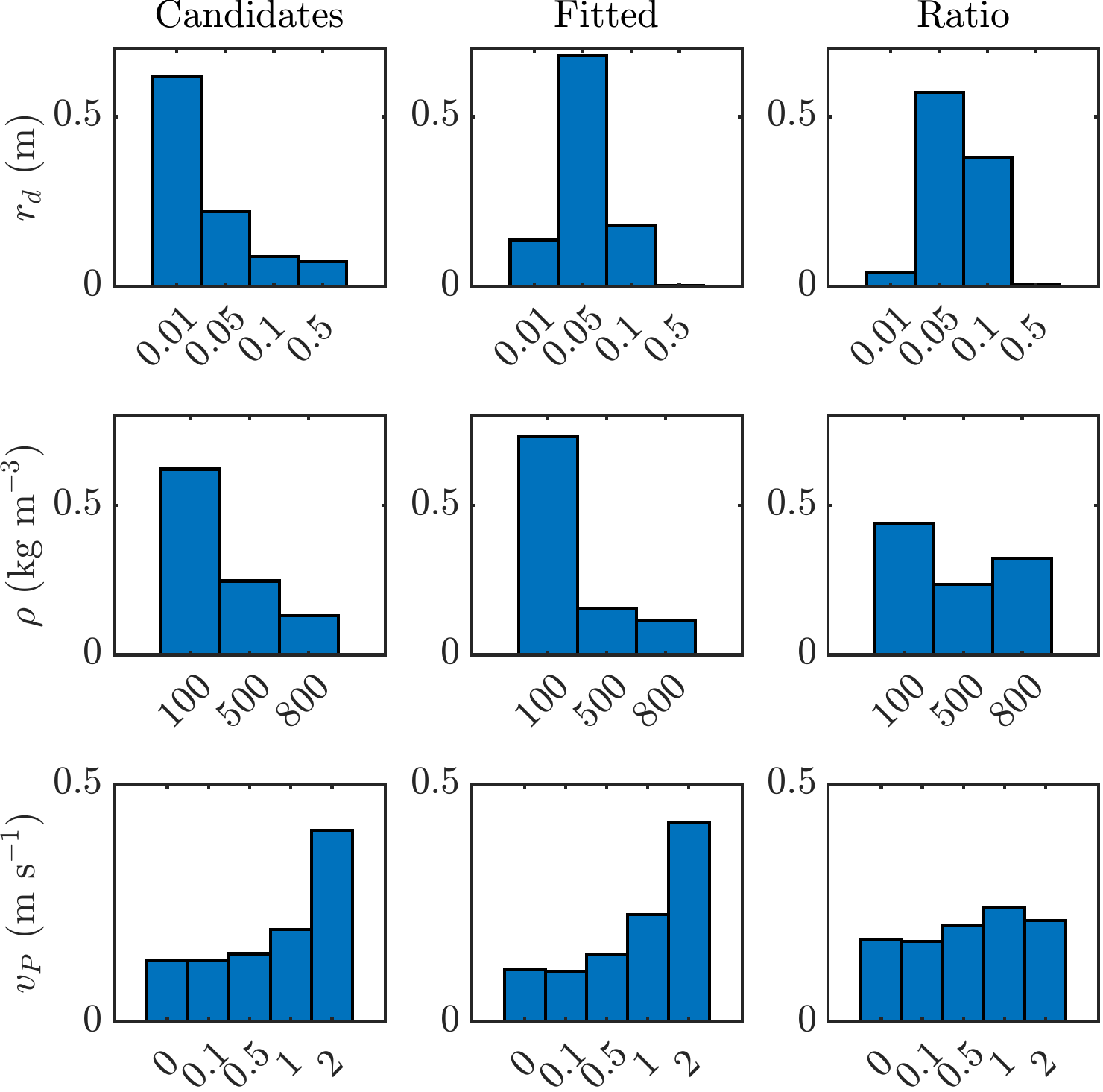}
        \caption{Distribution of physical properties of the aggregates for STP070.}
        \label{fig:distProp_70}
    \end{figure}
    \begin{figure}[ht]
        \centering
        \includegraphics[width=\linewidth]{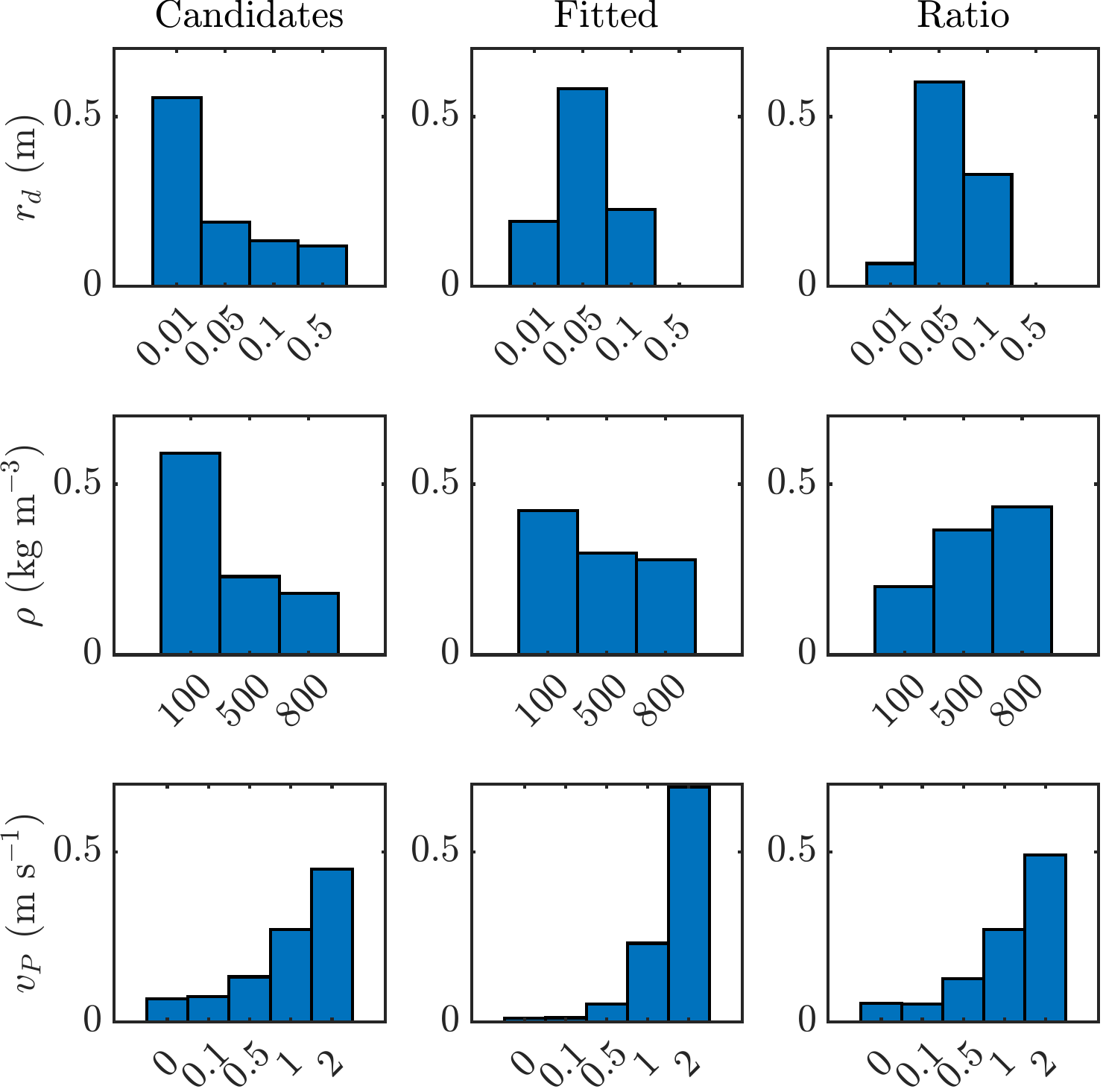}
        \caption{Distribution of physical properties of the aggregates for STP086.}
        \label{fig:distProp_86}
    \end{figure}
    \begin{figure}[ht]
        \centering
        \includegraphics[width=\linewidth]{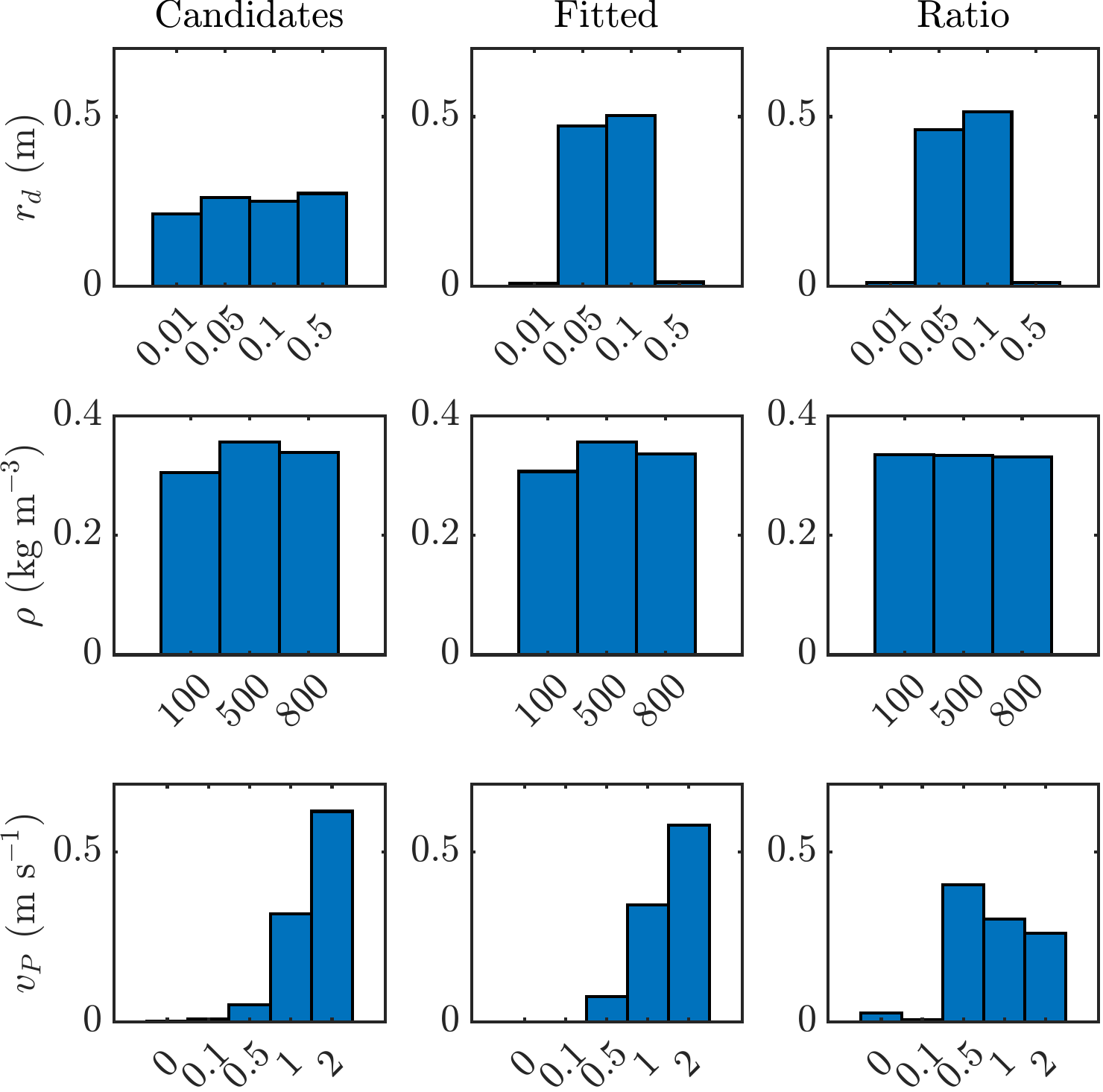}
        \caption{Distribution of physical properties of the aggregates for STP096.}
        \label{fig:distProp_96}
    \end{figure}

    \section{Source regions}\label{ap:sourceReg}
    \begin{figure*}[ht]
        \centering
        \includegraphics[width=.7\linewidth]{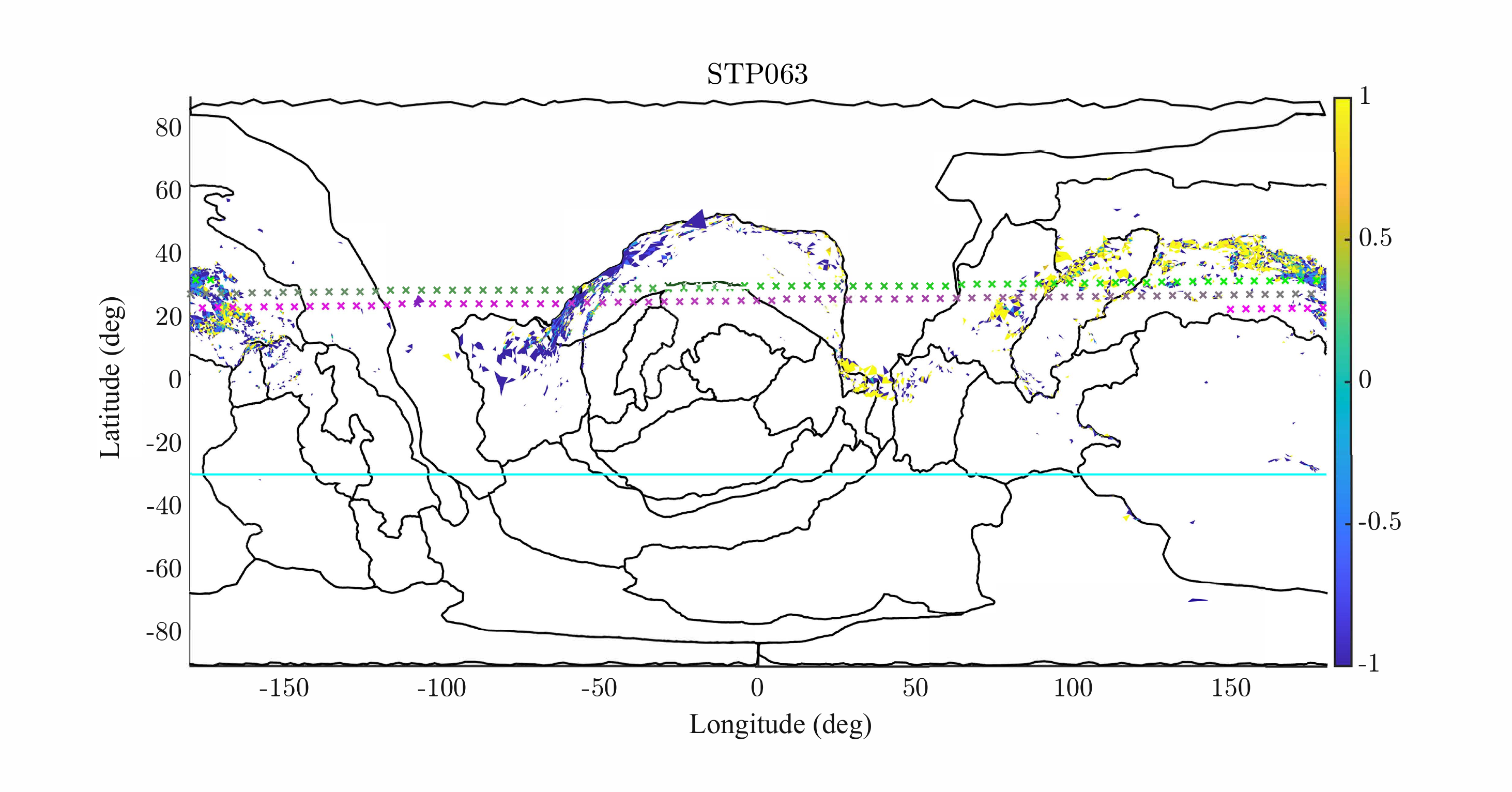}
        \caption{Ejection efficiency for image set STP063. Crosses indicate the projected spacecraft position from 12 hours before the start of the acquisition (green) to 12 hours after the end of it (magenta). Cyan solid line indicates the sub-solar latitude.}
        \label{fig:sourceReg_63}
    \end{figure*}
    \begin{figure*}[ht]
        \centering
        \includegraphics[width=.7\linewidth]{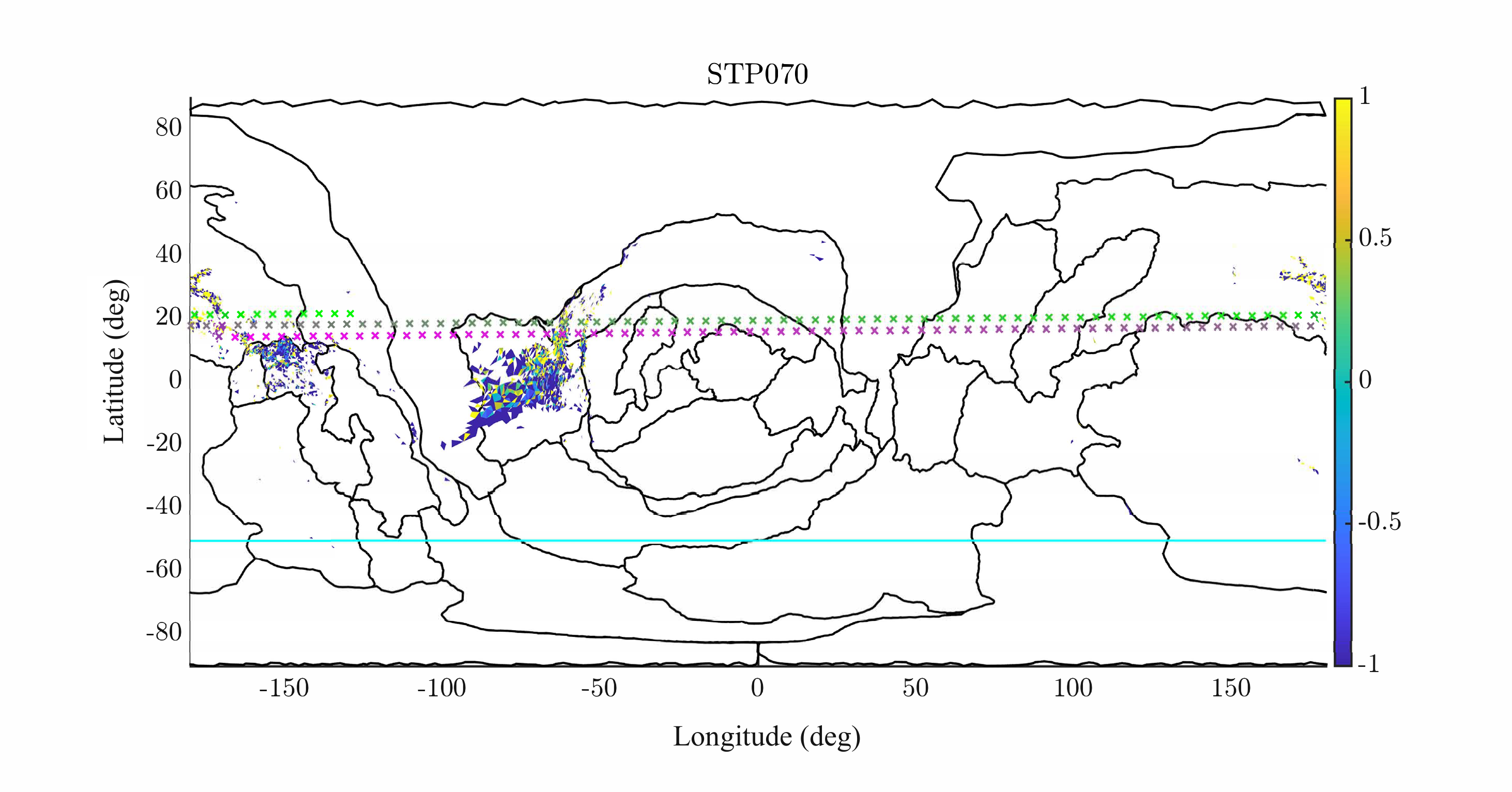}
        \caption{Ejection efficiency for image set STP070.}
        \label{fig:sourceReg_70}
    \end{figure*}
    \begin{figure*}[ht]
        \centering
        \includegraphics[width=.7\linewidth]{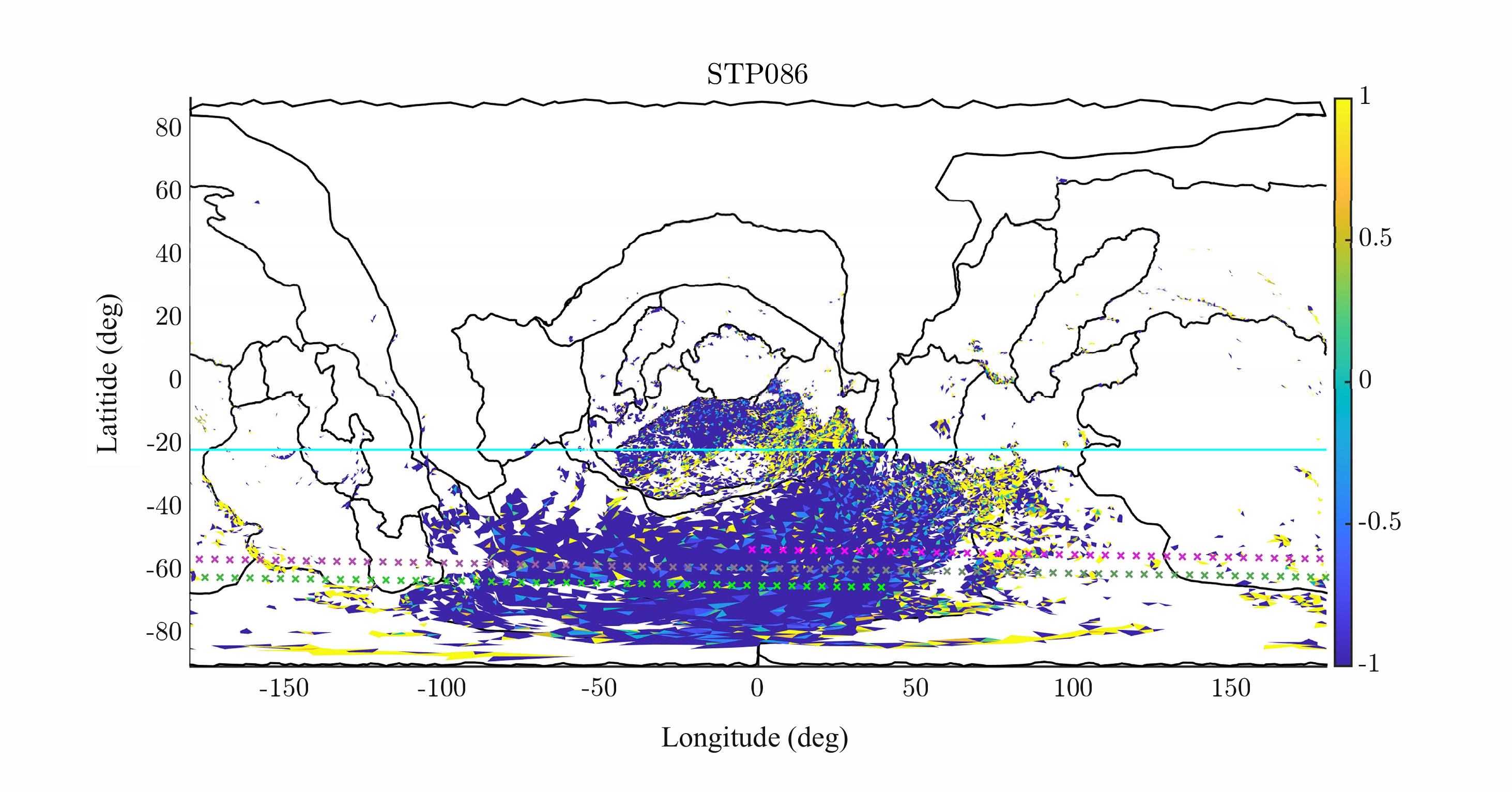}
        \caption{Ejection efficiency for image set STP086.}
        \label{fig:sourceReg_86}
    \end{figure*}
    \begin{figure*}[ht]
        \centering
        \includegraphics[width=.7\linewidth]{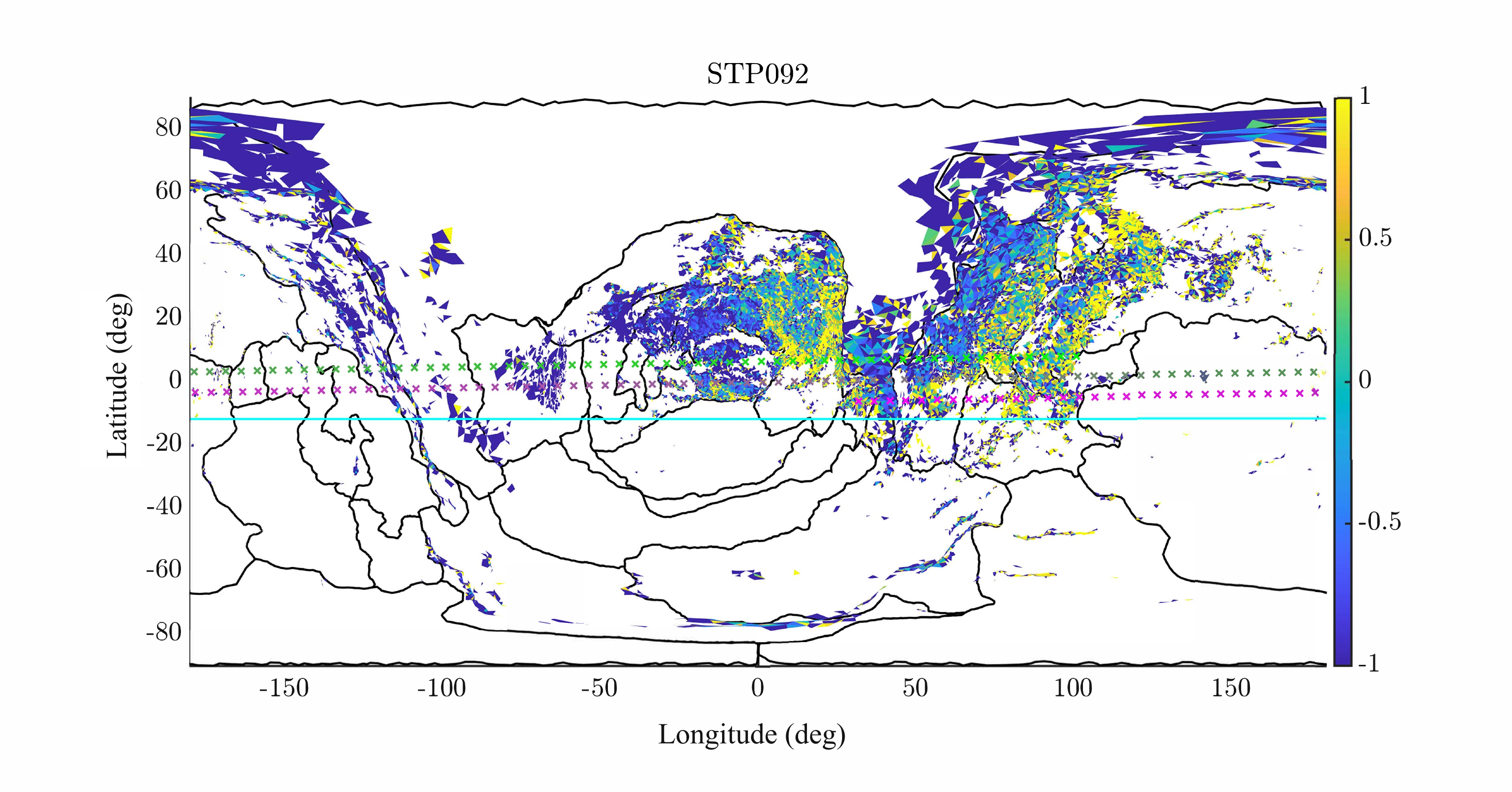}
        \caption{Ejection efficiency for image set STP092.}
        \label{fig:sourceReg_92}
    \end{figure*}
    \begin{figure*}[ht]
        \centering
        \includegraphics[width=.7\linewidth]{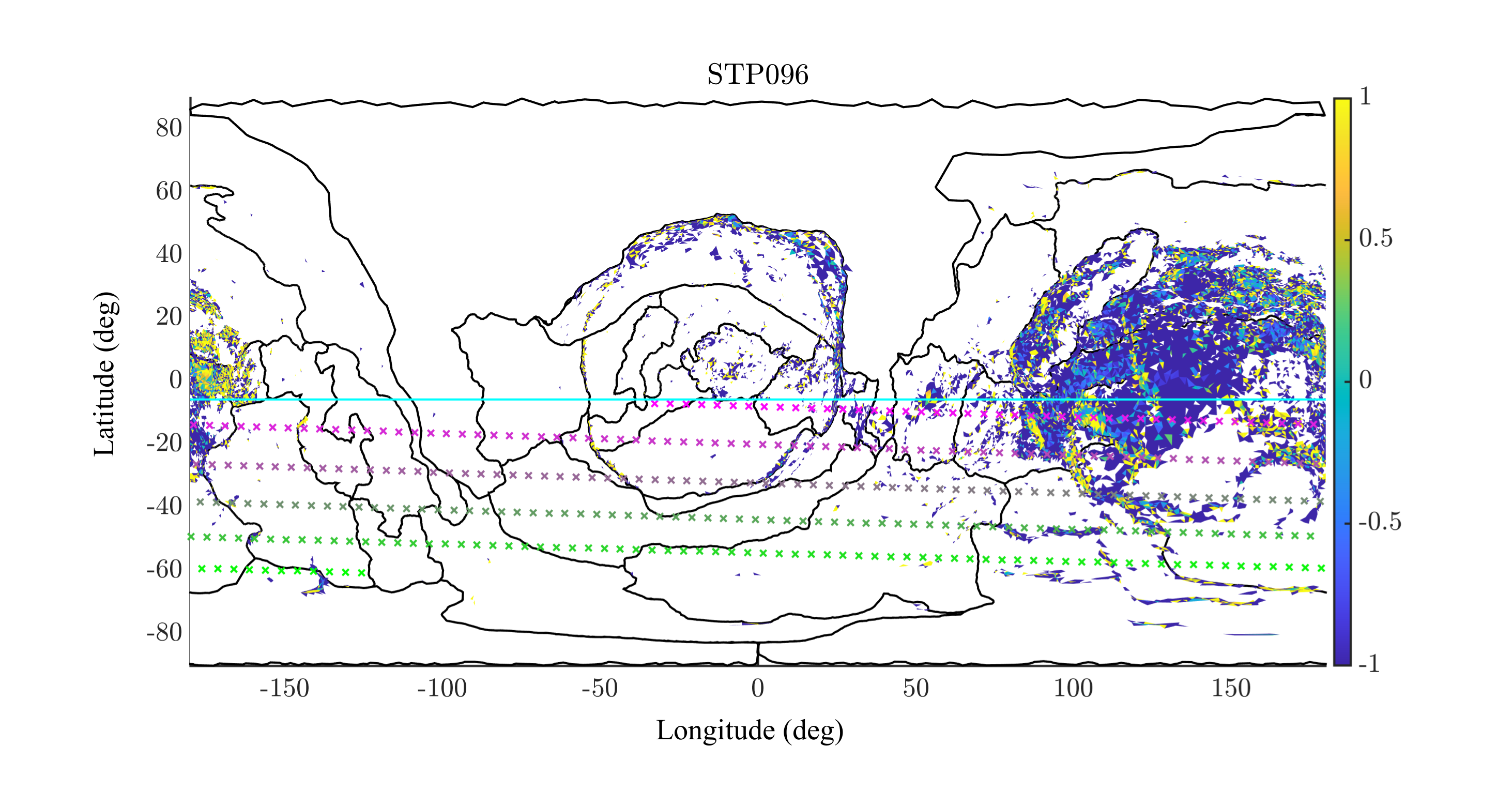}
        \caption{Ejection efficiency for image set STP096.}
        \label{fig:sourceReg_96}
    \end{figure*}

    \section{Corrected number of tracks per image}
    \begin{figure*}[ht]
        \centering
        \includegraphics[width=.45\linewidth]{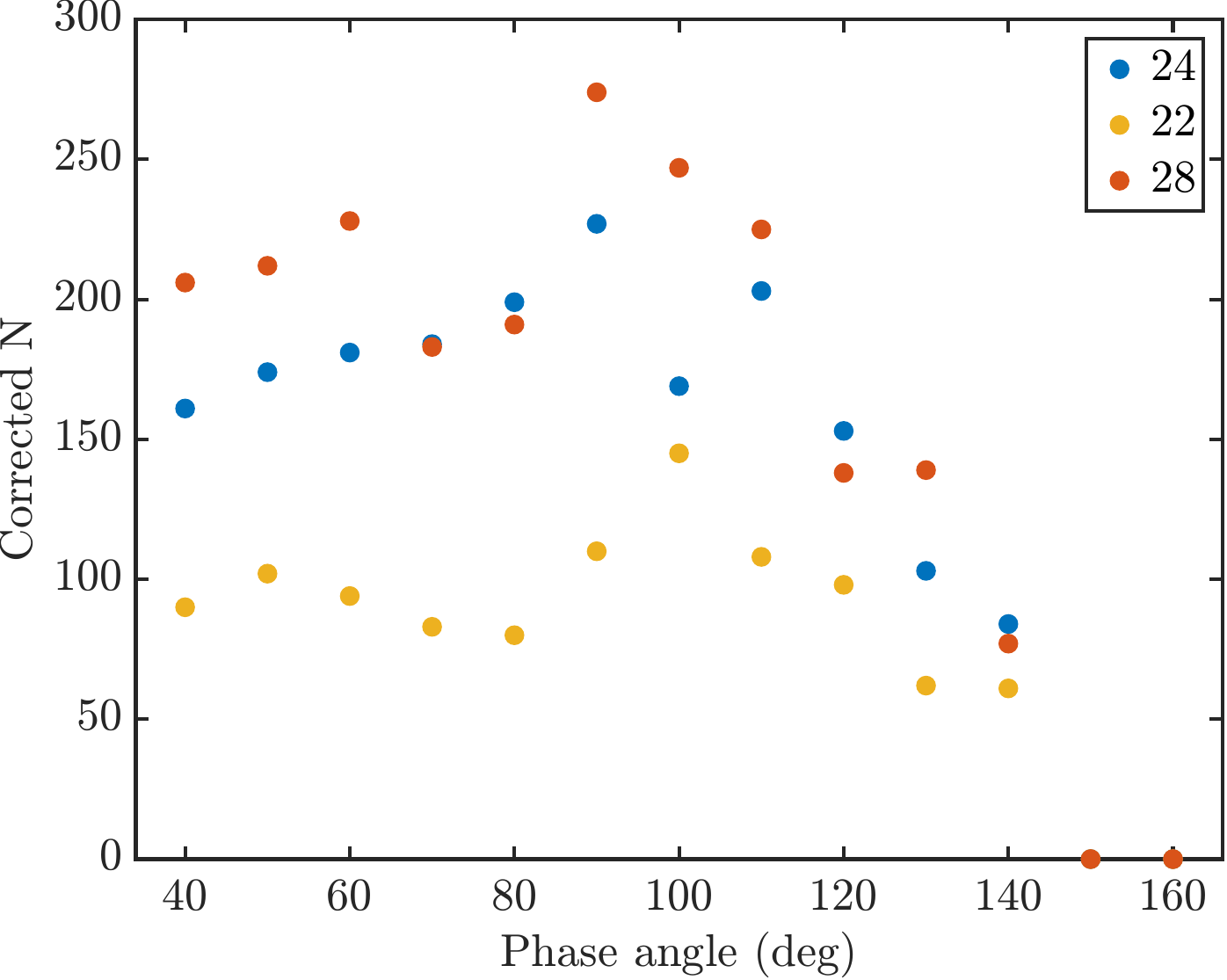}
        \includegraphics[width=.45\linewidth]{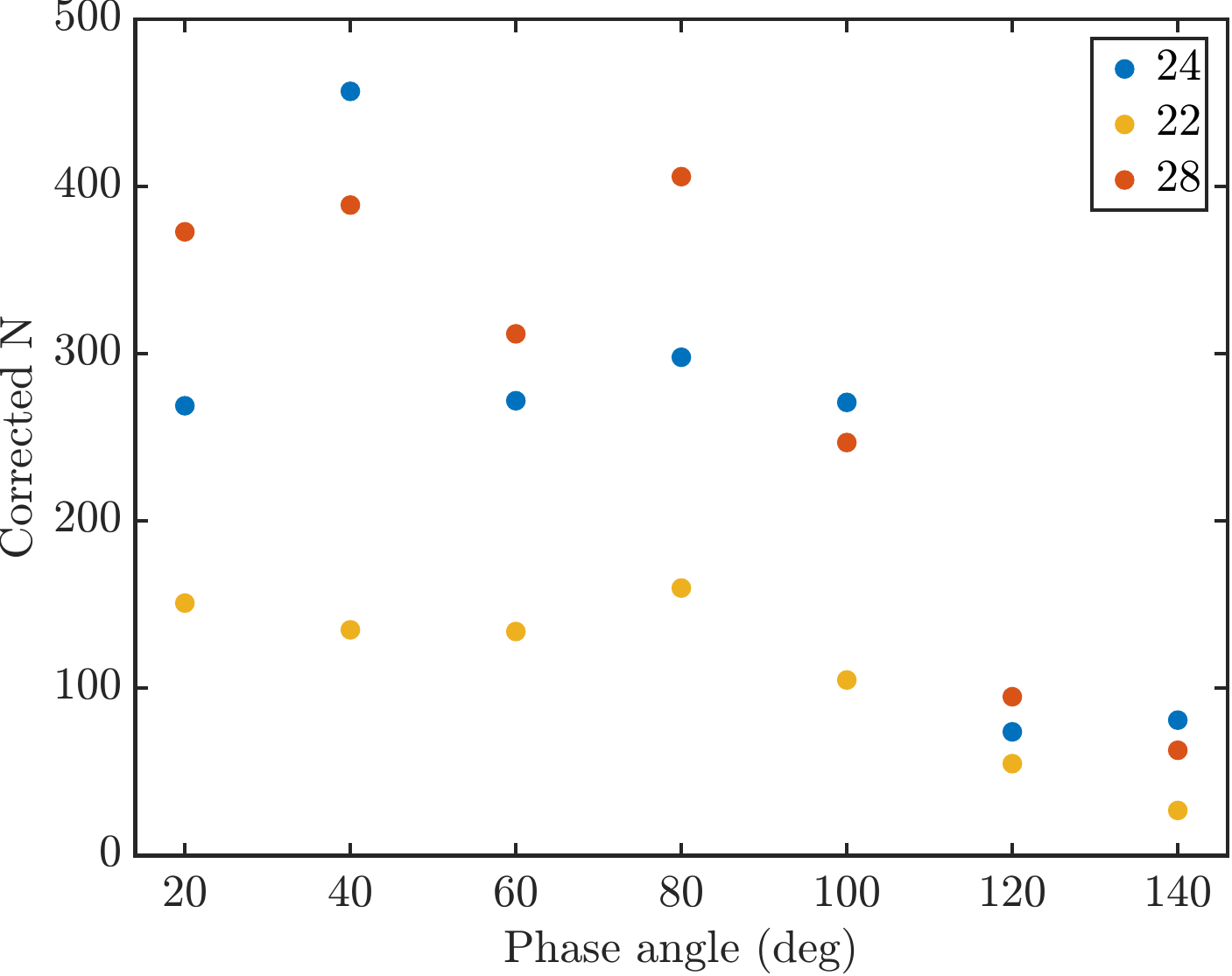}
        \includegraphics[width=.45\linewidth]{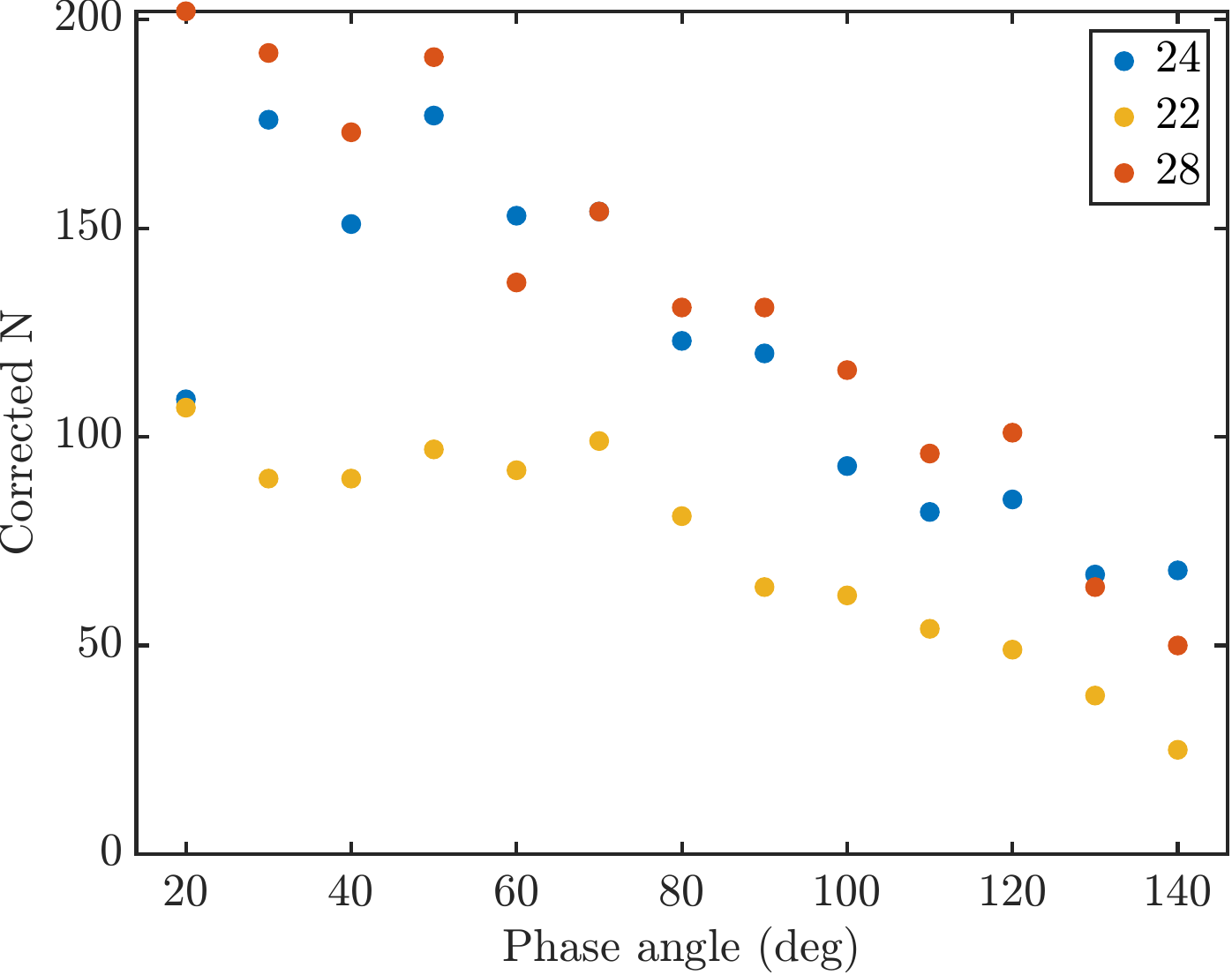}
        \includegraphics[width=.45\linewidth]{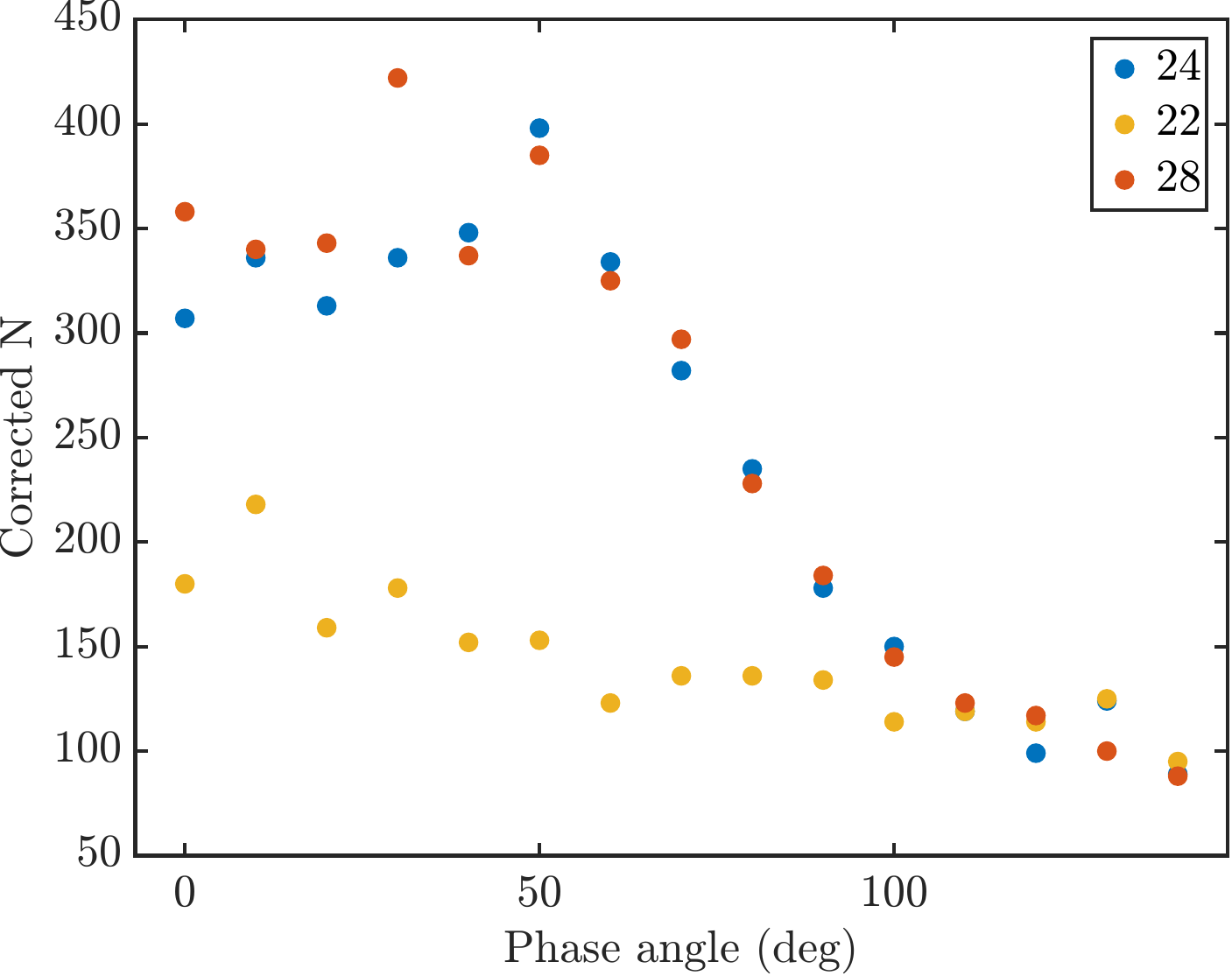}
        \caption{Corrected number of tracks per image. From top left are represented STP070, STP086, STP092 and STP096 respectively.}
    \end{figure*}
    
\end{appendix}

\end{document}